%% file: Main.tex
\documentclass[conference,compsoc]{IEEEtran}
\IEEEoverridecommandlockouts

\usepackage[hidelinks]{hyperref}
\usepackage{cite}
\usepackage{amsmath,amssymb,amsfonts}
\usepackage{algorithmic}
\usepackage{textcomp}
\def\BibTeX{{\rm B\kern-.05em{\sc i\kern-.025em b}\kern-.08em
    T\kern-.1667em\lower.7ex\hbox{E}\kern-.125emX}}

\usepackage{multirow}
\usepackage{graphicx}
\usepackage[table,xcdraw]{xcolor}
\usepackage{url}

\usepackage{pifont}
\newcommand{\ccmark}{\ding{52}}%
\newcommand{\xmark}{\ding{55}}%

\usepackage{listings}
\usepackage{tcolorbox}

\colorlet{punct}{red!60!black}
\definecolor{background}{HTML}{EEEEEE}
\definecolor{delim}{RGB}{20,105,176}
\definecolor{codegray}{rgb}{0.5,0.5,0.5}
\colorlet{numb}{magenta!60!black}

\lstdefinelanguage{json}{
    basicstyle=\normalfont\ttfamily,
    numberstyle=\scriptsize,
    stepnumber=1,
    numbersep=8pt,
    showstringspaces=false,
    breaklines=false,
    frame=single,
    backgroundcolor=\color{background},
    moredelim=[is][\color{black}]{<<}{>>},
    moredelim=[is][\color{codegray}]{<-}{->},
}

\usepackage{subcaption}

\newtheorem{definition}{Definition}

\usepackage[acronym,shortcuts]{glossaries-extra}
\glssetcategoryattribute{acronym}{nohyper}{true}
\setabbreviationstyle[acronym]{long-short}

\newacronym{acl}{ACL}{Access Control List}
\newacronym{ca}{CA}{Certificate Authority}
\newacronym{caam}{CAAM}{Cryptographic Acceleration and Assurance Modules}
\newacronym{cms}{CMS}{Cryptographic Message Syntax}
\newacronym{dos}{DoS}{Denial-of-Service}
\newacronym{dc}{DC}{Digital Certificate}
\newacronym{dhcp}{DHCP}{Dynamic Host Configuration Protocol}
\newacronym{id}{\(Dev_{ID}\)}{\(Dev_{ID}\)}
\newacronym{ec}{EC}{Elliptic Curve}
\newacronym{ecdl}{ECDL}{Elliptic Curve Discrete Logarithm}
\newacronym{eui}{EUI}{Extended Unique Identifier}
\newacronym{ietf}{IETF}{Internet Engineering Task Force}
\newacronym{iot}{IoT}{Internet-of-Things}
\newacronym{ids}{IDS}{Intrusion Detection System}
\newacronym{iiot}{IIoT}{Industrial IoT}
\newacronym{lldp}{LLDP}{Link Layer Discovery Protocol}
\newacronym{mmm}{\(MMM\)}{MUD Management Model}
\newacronym{mud}{MUD}{Manufacturer Usage Description}
\newacronym{mudb}{\(MUD\)-\(Binding\)}{Device-to-MUD Profile Binding}
\newacronym{mud-c}{\(Controller\)}{MUD Controller}
\newacronym{mud-class}{\(Class\)}{MUD Class}
\newacronym{mud-profile}{\(Profile\)}{MUD Profile}
\newacronym{mud-retrieval}{\(MUD\) \(Retrieval\)}{MUD File Retrieval}
\newacronym{mud-server}{\(Server\)}{MUD Server}
\newacronym{mud-table}{\(Table\)}{MUD Table}
\newacronym{mud-url}{MUD URL}{\ac{mud} \ac{url}}
\newacronym{mud-id}{MUD-\textit{ID}}{MUD Identifier}
\newacronym{nist}{NIST}{National Institute of Standards and Technology}
\newacronym{ok}{\textit{Operating Key}}{Operating Key}
\newacronym{os}{\textit{Operating Server}}{Operating Server}
\newacronym{oui}{OUI}{Organization Unique Identifier}
\newacronym{poc}{PoC}{Proof-of-Concept}
\newacronym{pki}{PKI}{Public Key Infrastructure}
\newacronym{se}{SE}{Secure Element}
\newacronym{tcp}{TCP}{Transmission Control Protocol}
\newacronym{tee}{TEE}{Trusted Execution Environment}
\newacronym{url}{URL}{Universal Resource Locator}
\newacronym{yang}{YANG}{Yet Another Next Generation}
\newacronym{zkp}{ZKP}{Zero-Knowledge Proof}

\usepackage{todonotes}

\newcommand{\name}{FIDEM}

\begin{document}

\title{FIDEM: A Standard-Compliant Framework for\\Secure Binding of MUD Profiles to IoT Devices}

\author{
\IEEEauthorblockN{Alessandro Lotto}
\IEEEauthorblockA{
\textit{University of Padua}\\ Padua, Italy \\ alessandro.lotto@math.unipd.it}
\and
\IEEEauthorblockN{Savio Sciancalepore}
\IEEEauthorblockA{
\textit{Eindhoven University of Technology}\\ Eindhoven, Netherlands \\ s.sciancalepore@tue.nl}
\and
\IEEEauthorblockN{Alessandro Brighente}
\IEEEauthorblockA{
\textit{University of Padua}\\
Padua, Italy \\
alessandro.brighente@unipd.it}
\and
\IEEEauthorblockN{Mauro Conti}
\IEEEauthorblockA{
\textit{University of Padua}\\
Padua, Italy \\
mauro.conti@unipd.it}
}


\maketitle

\begin{abstract}
The Manufacturer Usage Description (MUD) enables enforcement of network restrictions for IoT devices based on their expected network traffic, as specified by manufacturers in a MUD file.
Devices advertise a URL pointing to this file, yet the standard does not define how to securely bind the issuing device to its profile.
As a result, malicious devices can manipulate network policy enforcement by advertising valid URLs referencing genuine MUD profiles, but not intended for that device.
Although MUD defines a certificate-based secure issuance method, current deployments rely on the insecure DHCP-based extension due to simpler integration.
Existing solutions either depend on \ac{pki}, break standard compliance, require excessive active manufacturer involvement, or overlook secure profile updates.

In this paper, we present FIDEM, a standard-compliant framework for securing DHCP-based MUD URL issuance.
FIDEM provides cryptographic binding between IoT devices and their MUD profiles by leveraging Zero-Knowledge-Proof authentication, without requiring device certificates or device-side PKI, minimizing manufacturers' involvement, and supporting secure profile updates.
Formal analysis shows that FIDEM withstands stronger adversaries than in prior work, including supply-chain compromise and attacks using legitimate devices as cryptographic oracles.
Our real-world evaluation on two reference constrained devices (ESP32-S3 and ESP32-C6) demonstrates minimal overhead compared to standard DHCP (\(\sim 5 ms\), \(20 mJ\)) and significant improvements over certificate-based benchmarks (\(\times 21\) faster, \(\sim 23\%\) less energy consumption) on an ESP32-C6 device.
\end{abstract}


\input{Sections/1-Introduction}
\input{Sections/2-MUD}
\input{Sections/3-RelatedWork}
\input{Sections/4-SystemThreatModel}
\input{Sections/5-FIDEM_modified}
\input{Sections/6-SecurityAnalysis_modified}
\input{Sections/7-ExperimentalEvaluation}
\input{Sections/8-Extensions}
\input{Sections/9-Conclusions}


\newpage

\bibliographystyle{IEEEtran}
\bibliography{Bibliography}

\newpage
\appendices

\input{Appendix/Appendix}
\end{document}

%% file: Sections/1-Introduction.tex
\section{Introduction}  \label{sec:Introduction}
The increasing integration of \ac{iot} devices into cyber-physical systems is enhancing automation, process optimization, and safety~\cite{CybersecFor4.0, AddressingIndustry4.0}.
However, the constraints of such devices and networks make it challenging to deploy advanced resource-consuming security mechanisms~\cite{StudyIoT,LightweightIoT,LightIoT,ECCiot}.
Thus, \ac{iot} devices represent attractive targets for attackers seeking to exploit weak protections to access sensitive data or launch distributed attacks~\cite{IoTsecurity,MUDthread,Mirai}.
To mitigate this expanded attack surface, the \ac{ietf} recently introduced the \ac{mud} standard~\cite{MUD-rfc}.
\ac{mud} enables manufacturers to disclose the expected network behavior required for the proper functioning of their \ac{iot} devices, i.e., the \textit{MUD profile}\cite{DefBehavior}.
These profiles are formalized into \textit{\ac{mud} files}, which are signed by the manufacturer and stored on a \textit{MUD Server}.
A dedicated \textit{MUD Controller} to be deployed on the user side can retrieve the \ac{mud} file referenced by the \textit{\ac{mud} URL} issued by the \ac{iot} device at network joining time.
Then, the Controller can translate the \ac{mud} file into network policies and restrictions where any behavior not described in the file is considered unauthorized~\cite{MUD-rfc,DefBehavior,RoleDevice}.

Despite growing interest, several open problems in management, security, and implementation limit the real-world adoption of \ac{mud} today~\cite{DefBehavior,SoKMUD,RoleDevice,MUDscope,GatewayMUD}.
A critical concern is the \ac{mudb} vulnerability, which affects the security of the standard \ac{dhcp}-based URL issuance procedure~\cite{EnforcingBehavioral, AcceptableURL}.
Given the public nature of \ac{mud} URLs, a compromised device can issue a spoofed URL that points to an incorrect, yet authentic, \ac{mud} profile, thereby leading to the enforcement of incorrect network restrictions~\cite{EnforcingBehavioral, RoleDevice}.
The key factor enabling such attacks is that \ac{mud} decouples policy retrieval from device authentication.
The Controller accepts a \ac{mud} URL based on the authenticity of the retrieved file rather than any proof that the requesting device is entitled to use that profile~\cite{MUD-rfc, SoKMUD}.
Thus, a spoofed \ac{mud} URL does not inherently raise inconsistencies detectable by the Controller, and any device capable of advertising a valid URL can trigger the installation of (possibly incorrect) corresponding policies.
An extension based on X.509 is available for secure URL issuance~\cite{MUD-rfc}.
However, it is not supported in practice and current \ac{mud} implementations rather rely on the insecure \ac{dhcp} extension, due to more convenient plug-and-play device installation~\cite{CiscoMUD, NistMUD, osMUD, SoftMUD, DefBehavior, SoKMUD}.
At the same time, symmetric-key alternatives raise security concerns. 
The Controller would require access to device-specific secret keys, but an attacker could set up a malicious Controller to recover secret keys of target devices and reuse them to impersonate other devices.
Furthermore,
the solutions proposed in the literature present additional limitations, such as lack of standard compliance and secure \ac{mud} profile update mechanisms, or excessive manufacturer involvement during binding verification.
Such operational challenges make solving the \ac{mudb} vulnerability particularly challenging, and motivate the Research Question (RQ) driving this work:
\begin{tcolorbox}[colback=gray!10, colframe=gray!60!black, boxrule=0.5pt, arc=2pt, left=4pt, right=4pt, top=2pt, bottom=2pt]
\textbf{RQ}: How can we enforce cryptographically secure \ac{mudb} without requiring device certificates or device-side \ac{pki}, while minimizing active manufacturer involvement during verification?
\end{tcolorbox}

\textbf{Contribution}. In this paper, we introduce \name, a standard-compliant framework for secure \ac{dhcp}-based MUD URL issuance. \name\ establishes a cryptographic binding between the issuing device and the retrieved \ac{mud} profile leveraging Schnorr-based \ac{zkp} authentication, without requiring device certificates or device-side PKI for the binding operation, and minimizing manufacturer involvement.
We design \name\ to withstand a stronger, yet realistic threat model compared to prior work, explicitly capturing supply chain compromise scenarios and adversaries capable of exploiting legitimate devices as cryptographic oracles in place of compromised devices.
Furthermore, \name\ also introduces a unique formal \ac{mmm}, which explicitly specifies how manufacturers could efficiently maintain and associate \ac{mud} profiles within the system model.
We formally evaluate \name\ security properties with ProVerif and evaluate its performance against baseline DHCP-, HTTP- and TLS-based approaches in a \ac{poc} implementation with commercial \ac{iot} devices (ESP32-S3 and ESP32-C6).
Our results show minimal overhead compared to the insecure \ac{dhcp} extension (\(\sim 7\%, 5 ms\)), and significant savings compared to certificate-based approaches, i.e., \name\ is up to \(\times 21\) times faster and \(\sim 23\%\) less energy-consuming compared to the X.509 extension with TLS on a ESP32-C6 device.
In summary, this paper makes the following contributions:
\begin{itemize}
    \item We formalize a stronger yet realistic threat model for \ac{mud} than prior work, which also considers supply-chain compromise and adversaries capable of exploiting legitimate devices as cryptographic oracles.
    \item We present \name, a standard-compliant framework that enforces cryptographic binding between \ac{iot} devices and \ac{mud} profiles via lightweight \ac{zkp}.
    Compared to prior work, \name\ avoids device certificates and device-side \ac{pki} for \ac{mudb}, minimizes manufacturer involvement, and supports secure profile updates.
    \item We introduce a unique class-based MUD Management Model enabling scalable profile association, efficient key management, and lifecycle support.
    \item We implement \name\ on real \ac{iot} devices, compare it to state-of-the-art benchmark approaches, and make the source code available open source~\footnote{\url{https://anonymous.4open.science/r/FIDEM-B103/README.md}} (anonymized for submission).
\end{itemize}

\textbf{Roadmap}. The remainder of this paper is organized as follows.
Sec.~\ref{sec:MUD} introduces the \ac{mud} standard. 
Sec.~\ref{sec:RelatedWork} reviews vulnerabilities and limitations in existing solutions.
Sec.~\ref{sec:SysModel} outlines our system and threat model.
Sec.~\ref{sec:Fidem} provides the details of \name.
Sec.~\ref{sec:SecurityAnalysis} presents a discussion on its security guarantees.
Sec.~\ref{sec:Evaluation} reports experimental results.
Sec.~\ref{sec:Limitations} discusses limitations and future directions.
Finally, Sec.~\ref{sec:Conclusions} concludes the paper.

%% file: Sections/2-MUD.tex
\section{Preliminaries on \ac{mud}} \label{sec:MUD}
\ac{mud} grounds on the observation that most \ac{iot} devices have a narrow operational scope and, therefore, require only a restricted set of network interactions~\cite{RoleDevice}.
To enforce this principle, manufacturers formally define the expected network behavior of their \ac{iot} devices through a set of \acp{acl}.
These specify the minimal set of necessary and allowed IP-based network communications for the correct functioning of the \ac{iot} device~\cite{SoftMUD}.
The collection of the \acp{acl} constitutes the device’s MUD profile, which is expressed in a \textit{\ac{mud} file} using \ac{yang} modules and serialized in JSON format~\cite{eMUD, MUD-rfc}.
Thus, the \ac{mud} file serves as a formal declaration of the device’s native operational requirements. 

\begin{figure}[!t]
    \centering
    \includegraphics[width=\columnwidth]{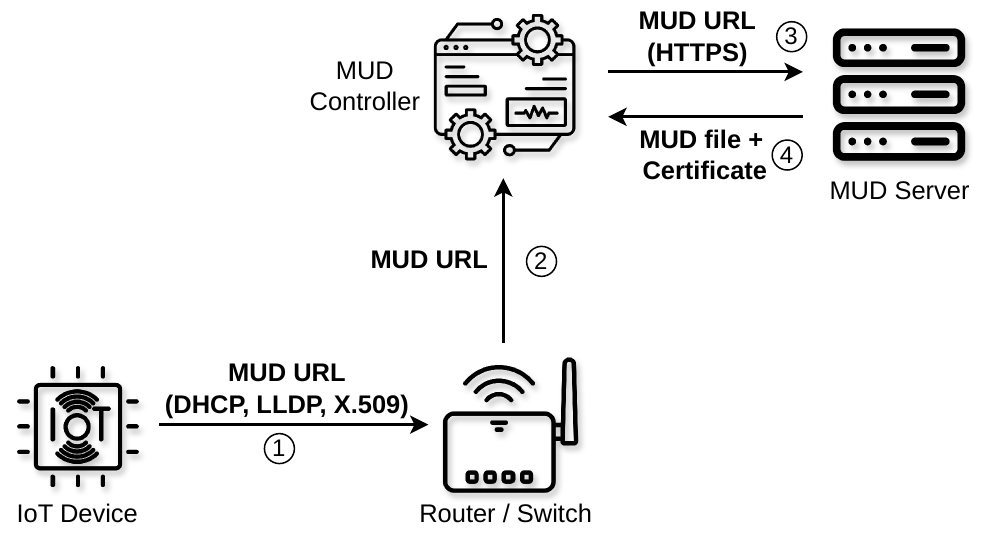}
    \caption{\ac{mud} architecture and file retrieval procedure~\cite{MUD-rfc}.}
    \label{fig:MUDarchitecture}
\end{figure}

\textbf{Architecture and File Retrieval.}
Fig.~\ref{fig:MUDarchitecture} illustrates the \ac{mud} architecture and file retrieval procedure~\cite{MUD-rfc}.
The process begins when the \ac{iot} device issues a \textit{\ac{mud} \ac{url}} to the network router (step 1), which identifies the location of the device’s \ac{mud} file.
The router forwards the \ac{url} to the MUD Controller (step 2), which in turn retrieves the \ac{mud} file from the MUD Server via HTTPS (step 3).
Before applying the rules, the MUD Controller verifies the profile authenticity using the associated \ac{cms} object (step 4).
Once validated, the MUD Controller translates the \acp{acl} specified in the file into enforceable network policies, which the router applies to restrict the device’s communications.
According to the standard, the MUD profile remains valid as long as the device is connected to the network~\cite{MUD-rfc}.
In some deployment scenarios, such as smart homes, the MUD Controller logic can be integrated directly into the router, avoiding \ac{url} relaying~\cite{eMUD, RoleDevice, NistMUD}.

\textbf{URL Issuance}.
The \ac{mud} standard defines three extension for URL issuance.~\cite{MUD-rfc}.
The X.509 extension introduces a non-critical certificate field compliant with IEEE 802.1AR, that contains a single \ac{url} pointing to the online \ac{mud} file.
In this case, the legitimacy of the \ac{mud} \ac{url} relies on the authenticity of the certificate used for device authentication.
Conversely, the \ac{dhcp} and \ac{lldp} extensions embed the \ac{url} into a string \(\texttt{MUDstring} = \texttt{MUDurl} \,,\, \left[^"\,\,\,^" \,,\, \texttt{reserved}\right]\), where \texttt{MUDurl} is an HTTP-based \ac{url} compliant with RFC7230~\cite{rfc7230}, and \texttt{reserved} is reserved for future extensions.
Unlike X.509, \ac{dhcp} and \ac{lldp} do not provide any built-in mechanism to authenticate the transmitted \ac{url} and ensure \ac{mudb}.
This lack of protection introduces security issues, such as spoofed \ac{mud} advertisements, which can lead to incorrect enforcement of overly permissive profiles.
Such misconfigurations can expose the network to compromise, enable lateral movement, and facilitate privilege escalation attacks~\cite{RoleDevice, AcceptableURL}.

%% file: Sections/3-RelatedWork.tex
\section{Related Work}  \label{sec:RelatedWork}
The MUD URL issuance mechanism for \ac{mud} retrieval is the cornerstone of the \ac{mud}, determining which set of network policies (i.e., MUD file) the Controller should enforce based on the received \ac{url}.
\ac{dhcp} and \ac{lldp} extensions lack security guarantees to ensure the integrity of the issued \ac{url}~\cite{MUD-rfc, DefBehavior}.
To the best of our knowledge, no prior work focuses on securing these extensions, although \ac{dhcp} is extensively considered in \ac{mud} implementations and other works using \ac{mud}~\cite{MitigatingMUD, NistMUD, chang2018proactive, SoftMUD, RoleDevice}.
Table~\ref{tab:Comparison} summarizes existing approaches specifically designed to secure the \ac{mud} URL issuance mechanism.

\input{Tables/Comparison}

\textbf{Device Authentication vs. MUD-Binding}.
Solutions such as eMUD~\cite{eMUD}, MUDscan~\cite{MUDscan}, and MUDThread~\cite{MUDthread} authenticate devices before retrieving the \ac{mud} file.
However, eMUD is vulnerable to replay attacks as it relies on static \acp{dc} with no challenge-response interaction.
MUDscan introduces challenge-response authentication, but using the manufacturer’s public key.
Thus, it mainly ensures that the device belongs to the specific manufacturer, rather than providing \ac{mudb}.
Similarly, MUDThread and MeshGuard leverage IEEE 802.15.4 link layer encryption and authentication~\cite{sciancalepore2017_ett, meshguard}, but do not ensure \ac{mudb}.
Lastly, proposals from \textit{Jimenez J.}~\cite{MUDCoAP} and \textit{Richardson et al.}~\cite{MUDqr} do not provide any cryptographic mechanism to protect the issued \ac{url}.
These examples show that \ac{mudb} is a distinct security property than device authentication.
A compromised device can first authenticate using legitimate credentials, then issue a spoofed \ac{mud} URL.
Thus, \ac{mudb} is a challenging security property to ensure beyond device authentication.

\textbf{Manufacturer-Involved Approaches}.
\textit{Garcia et al.}~\cite{EnforcingBehavioral} and \textit{Matheu et al.}~\cite{SecurityArchitecture} adopt a different strategy for \ac{mud} file retrieval: the MUD Controller receives the \ac{url} directly from the manufacturer after successful device authentication.
While this strategy guarantees binding (assuming correct authentication), it introduces significant operational overhead for manufacturers.
In fact, they must maintain live authentication services, perform cryptographic operations, and handle additional traffic for \ac{mud} purposes.
We argue that such a workload primarily serves end users and rather discourages manufacturers from adopting \ac{mud}, thus limiting its widespread adoption.
To promote adoption, any solution for secure \ac{mudb} should minimize manufacturers involvement beyond generating, updating, and serving \ac{mud} files.

\textbf{\ac{pki} Limitations}.
A proposal from \textit{Richardson et al.}~\cite{AcceptableURL} offers secure binding and is fully compliant with the \ac{mud} standard, without requiring active manufacturer involvement during \ac{mud} file retrieval.
However, it relies on X.509 certificates and a \ac{pki}, which have seen limited support in practice.
Although RFC 8520 defines a non-critical X.509 extension, to the best of our knowledge, no available or open implementations currently support it~\cite{DefBehavior, NistMUD, CiscoMUD, SoftMUD, osMUD, SoKMUD}.
Furthermore, native X.509/802.1AR support remains limited in consumer \ac{iot} devices due to operational issues related to the certificate size, computational overhead, and lifecycle management challenges for constrained devices~\cite{LightIoT, LightweightIoT, LightweightX509, EnforcingBehavioral, SecurityArchitecture}.
This is also highlighted by the fact that the majority of related works avoid relying on \ac{pki} (see Table~\ref{tab:Comparison}).
Recent efforts, such as C509 certificates, aim to provide compact encodings for constrained environments~\cite{C.509}.
However, these are still in draft status and have not yet been standardized.
By contrast, \ac{dhcp} remains the \textit{de facto} baseline for IP address assignment in \ac{iot}~\cite{DHCPusage}, driving industry preference for \ac{dhcp} extension over X.509.
This trend creates a deployment-driven motivation to focus on strengthening the \ac{dhcp} extension.

\textbf{Standard Compliance.}
None of the previous works, except~\cite{AcceptableURL}, is standard-compliant (see Table~\ref{tab:Comparison}).
Breaking the standard compliance for \ac{mud} file retrieval creates interoperability issues and increases the complexity of management and enforcement.
In contrast, a standard-compliant solution is highly desirable, as it favors interoperability and adoption.

\textbf{Secure MUD Profile Update.}
A secure \ac{mud} profile update mechanism is crucial to maintain the integrity and consistency of enforced policies over time.
Without it, attackers could exploit outdated or tampered profiles, potentially bypassing security restrictions.
Despite its importance, existing solutions except \textit{Richardson et al.}~\cite{AcceptableURL} do not address this aspect, leaving a remarkable gap in the overall security of the \ac{mud} framework.

\textbf{Alternative Approaches}.
Alternative approaches, such as remote attestation or certificateless encryption, may seem viable solutions to ensure \ac{mudb}.
Remote attestation verifies that the device is in a genuine state~\cite{SurveyRemote, CollectiveAttestation, RATA}, which could indirectly support trust in the issued \ac{url}.
However, its scope is broader than \ac{mudb}, as it focuses on device integrity rather than the correctness of specific configuration data.
Moreover, secure attestation typically requires expensive hardware roots-of-trust and computational energy~\cite{SurveyRemote, RATA}.
On the other hand, certificateless encryption aims to reduce dependence on traditional \acp{dc}.
However, it still relies on a trusted key generation center and introduces key escrow concerns~\cite{GenericCertificateless, ReliableCertificateless, SurveyCertificateless}.
These limitations make both approaches impractical for lightweight, standard-compliant \ac{mud} deployments, where simplicity and interoperability are key factors.

In summary, existing solutions fail to guarantee the \ac{mudb} property, impose excessive operational overhead on manufacturers, or lack standard compliance.
Furthermore, existing proposals overlook secure profile updates, leaving devices vulnerable to outdated or tampered configurations.
Alternative approaches, such as remote attestation or certificateless encryption, introduce complexity and hardware requirements incompatible with constrained environments.
These gaps highlight the need for a lightweight, standard-compliant mechanism that enforces \ac{mudb} without relying on heavy cryptographic infrastructures or requiring active manufacturer involvement.

%% file: Tables/Comparison.tex
\begin{table*}[!t]
\renewcommand{\arraystretch}{1.25}
\centering
\caption{Comparison of existing approaches for \ac{mud} file retrieval and MUD-Binding verification.}
\label{tab:Comparison}
\resizebox{\textwidth}{!}{%
\begin{tabular}{l|c|c|c|c|c|c|c|c}
\rowcolor[HTML]{EFEFEF} 

\hline
\multicolumn{1}{c|}{\cellcolor[HTML]{EFEFEF}\textbf{Method}} &
\textbf{\begin{tabular}[c]{@{}c@{}}\ac{url} Issuing\\ Method\end{tabular}} &
\textbf{\begin{tabular}[c]{@{}c@{}}Cryptographic\\ Approach\end{tabular}} &
\textbf{\begin{tabular}[c]{@{}c@{}}MUD std\\ Compliance\end{tabular}} &
\textbf{\begin{tabular}[c]{@{}c@{}}Avoid \ac{pki}\\for MUD Binding\end{tabular}} &
\textbf{\begin{tabular}[c]{@{}c@{}}Define\\ MMM\end{tabular}} &
\textbf{\begin{tabular}[c]{@{}c@{}}Avoid Manufacturer\\ Involvement\end{tabular}} &
\textbf{\begin{tabular}[c]{@{}c@{}}Secure\\ URL Update\end{tabular}} &
\textbf{MUD-Binding} \\ \hline

\textit{eMUD}~\cite{eMUD} & Not specified  & Certificate-based Authentication & \xmark & \xmark  & \xmark  & \xmark  & \xmark  & \xmark  \\

\rowcolor[HTML]{EFEFEF} 
\textit{MUDscan}~\cite{MUDscan} & Not specified  & Public-Key challenge-response  & \xmark & \xmark & \xmark  & \xmark  &  \xmark & \xmark \\

\textit{MUDThread}~\cite{MUDthread} & \begin{tabular}[c]{@{}c@{}}Mesh Link Establishment \\ Parent Request \end{tabular} & \begin{tabular}[c]{@{}c@{}}AES-128 CCM\\(IEEE 802.15.4 link layer) \end{tabular} & \xmark & \ccmark & \xmark & \ccmark & \xmark & \xmark \\

\textit{MeshGuard}~\cite{meshguard} & \begin{tabular}[c]{@{}c@{}}Mesh Link Establishment \\ Parent Request \end{tabular} & \begin{tabular}[c]{@{}c@{}}AES-128 CCM\\(IEEE 802.15.4 link layer) \end{tabular} & \xmark & \ccmark & \xmark & \ccmark & \xmark & \xmark \\

\rowcolor[HTML]{EFEFEF} 
\textit{Garcia et al.}~\cite{EnforcingBehavioral}  & PANA  & EAP-based authentication & \xmark  & \ccmark & \xmark  & \xmark  & \xmark & \ccmark \\

\textit{Matheu et al.}~\cite{SecurityArchitecture}   & CoAP  & EAP-based authentication & \xmark & \ccmark & \xmark  & \xmark  & \xmark & \ccmark \\

\rowcolor[HTML]{EFEFEF} 
\textit{Richardson et al.}~\cite{AcceptableURL} &  X.509 & Certificate-based Authentication & \ccmark & \xmark & \xmark & \ccmark & \ccmark \(^{*}\) & \ccmark \(^{*}\) \\

\textit{Jimenez J.}~\cite{MUDCoAP} & \begin{tabular}[c]{@{}c@{}}CoAP \end{tabular} & \textit{N.A.} & \xmark & \ccmark & \xmark & \ccmark & \xmark & \xmark \\

\rowcolor[HTML]{EFEFEF} 
\textit{Richardson et al.}~\cite{MUDqr} & \begin{tabular}[c]{@{}c@{}}QR-codes \end{tabular} & \textit{N.A.} & \xmark & \ccmark & \xmark & \ccmark & \xmark & \xmark \\

\hline
\textit{FIDEM} & DHCP & EC-based ZKP authentication & \ccmark & \ccmark & \ccmark & \ccmark & \ccmark & \ccmark \\
\hline
\end{tabular}%
}

\footnotesize
\vspace{.25em}
    \begin{itemize}
        \item[] * \ac{mudb} assurance for profile updates relies on manufacturer's secure \acp{url} management.
    \end{itemize}
\end{table*}

%% file: Sections/4-SystemThreatModel.tex
\section{System and Threat Model}   \label{sec:SysModel}
\begin{figure}[!t]
    \centering
    \includegraphics[width=\columnwidth]{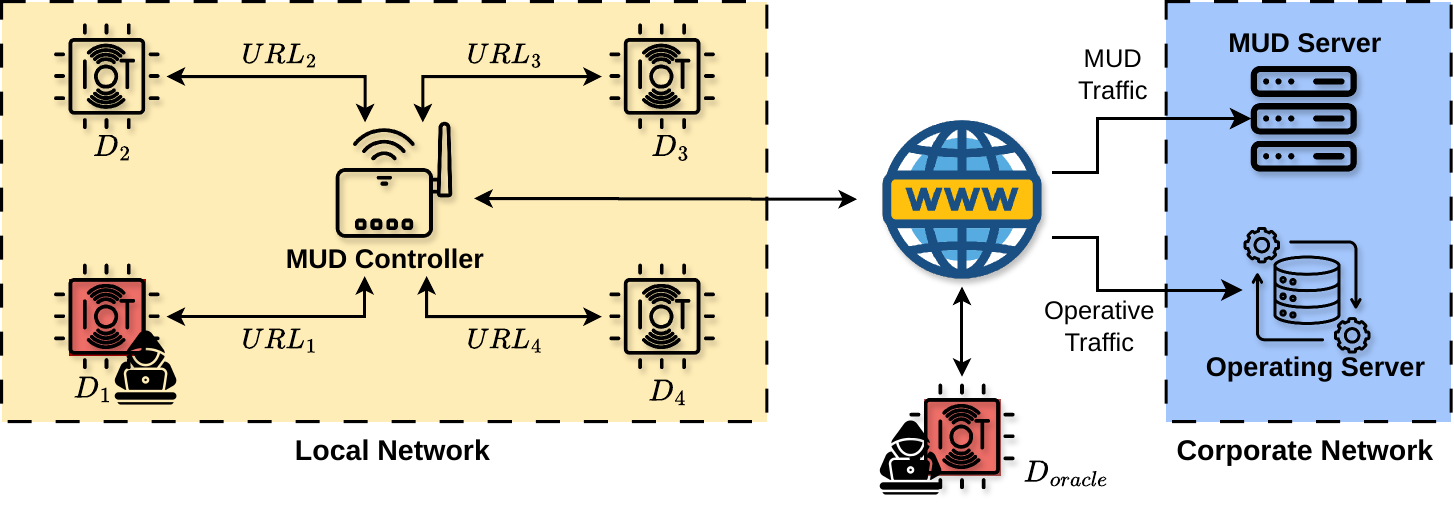}
    \caption{System and threat model. \(D_1\) is compromised, and the attacker may use legitimate device(s) \(D_{oracle}\) as a cryptographic oracle to perform operations on behalf of \(D_1\).}
    \label{fig:SystemModel}
\end{figure}

\textbf{System Model}.
Fig.~\ref{fig:SystemModel} illustrates the system model considered in this work.
We consider a local \ac{mud}-enabled \ac{iot} network composed of heterogeneous resource-constrained devices, produced by various manufacturers and connected to a central router.
Devices issue their \ac{mud} URL using the available \ac{dhcp} extension, in line with most literature~\cite{eMUD,MUDscan,SecurityArchitecture}.
Following common practice, we integrate the MUD Controller into the router~\cite{RoleDevice, NistMUD}.
To clearly delineate the scope of this work, we distinguish between the \emph{\ac{mud} Domain} and the \emph{Operational Domain}.
The former includes the MUD Controller, the MUD Server, and all traffic and operations governed by the \ac{mud} framework.
Conversely, the latter encompasses all existing infrastructure, services, and procedures required for device functionality, independent of \ac{mud}.
Within the \textit{Operational Domain}, we model an \textit{Operating Server} that handles routine device operations such as updates, authentication, and management tasks.
These are performed using pre-existing mechanisms and are outside the scope of \ac{mud} and this work.
By contrast and according to the limitations identified in Sec.~\ref{sec:RelatedWork}, we assume that the MUD Server is a passive entity whose sole role is to serve the \ac{mud} file corresponding to a given \ac{url}.
It does not participate in cryptographic operations related to authentication and \ac{mudb} verification.
Finally, we adopt a standard cryptographic model in which secret material stored on devices is protected in secure hardware.
Although we acknowledge that some \ac{iot} devices lack strong hardware protections, this assumption is consistent with a large body of prior work~\cite{MUDscan, hardware1, hardware2, hardware3, hardware4}.

\textbf{Threat Model.}
We consider a realistic and conservative threat model, drawing on examples and motivations from real life.
We consider that \ac{iot} devices cannot be assumed trustworthy at the time of deployment, as they can be compromised at any stage of their lifecycle, including manufacturing or distribution, reflecting well-documented exposure of \ac{iot} supply chains~\cite{IoTcompromise1, IoTcompromise2, IoTcompromise3}.
We thus focus on two representative compromise scenarios, namely \emph{Compromised Legitimate Device} and \emph{Counterfeit Device Impersonation}, both leveraging binding limitations of the \ac{mud} standard architecture~\cite{MUD-rfc}.
In the former, the attacker compromises a genuine device and aims to induce the enforcement of a more permissive network policy than the one originally intended for that device.
Since \ac{mud} URLs are publicly accessible, an attacker can systematically analyze the product ecosystem of a manufacturer identifying those with more permissive policies.
Compromised devices thus advertise spoofed URLs corresponding to other devices from the same manufacturer, whose \ac{mud} profile grants broader communication privileges.
This behavior is consistent with prior observations in the literature~\cite{eMUD,MUDscan}.
In the latter scenario, the attacker introduces counterfeit devices into the system, presenting them as legitimate products of a target manufacturer.
Although this may appear an overly strong adversary, it is in fact supported by real-world incidents involving fraudulent networking and \ac{iot} equipment commercialized through third-party vendors or online marketplaces~\cite{IoTcompromise1, IoTcompromise2, IoTcompromise3}.
In such a scenario, since the attacker does not need to break cryptographic protections or impersonate manufacturer signatures, counterfeit devices can simply advertise a valid URL associated with a legitimate \ac{mud} file.
Device-level authentication, if any, typically occurs at later stages (e.g., during Cloud-based onboarding), and it does not prevent such attack, leading to the enforcement of (incorrect) network policies based on the spoofed URL~\cite{SecurityArchitecture, SoKMUD, EnforcingBehavioral}.
In line with our system model, we consider an attacker that cannot directly access secret material stored on devices or the manufacturer's private keys.
However, the attacker can access to one or more legitimate devices of the target manufacturer.
This reflects a realistic capability, as such devices are commercially available and can be freely acquired.
Thus, the attacker can leverage them as cryptographic oracles, enabling cryptographic operations without possessing the underlying secrets.
We model the set of all such devices as a logical entity \(D_{oracle}\), which may be locally or remotely accessible.
Communication between a compromised device \(D_1\) and \(D_{oracle}\) is assumed to occur through the network infrastructure, in particular via the router enforcing \ac{mud} policies.
This interaction can be facilitated by permissive rules commonly observed in \ac{mud} profiles, such as those allowing communication among devices from the same manufacturer~\cite{RoleDevice, SoKMUD, MUDscan}.
These rules make such communication appear as legitimate from the network perspective, further complicating detection.
We explicitly exclude out-of-band channels (e.g., Bluetooth, cellular) from the model, to maintain focus on network-enforced policies.

\textbf{Attacker objectives}.
Within this threat model, the attacker aims to achieve three objectives: (i) recover secret material through passive observation of communications (\textbf{O1}), mislead the Controller into classifying a device as belonging to a different manufacturer (\textbf{O2}), and induce the enforcement of an unintended \ac{mud} profile within the same manufacturer domain (\textbf{O3}).
The latter captures the core \ac{mudb} vulnerability, namely, the ability of an attacker to exploit the lack of binding between devices and their declared \ac{mud} profile.
We analyze the security of \name\ against these objectives in Sec.~\ref{sec:SecurityAnalysis}.

%% file: Sections/5-FIDEM_modified.tex
\section{FIDEM} \label{sec:Fidem}
This section presents \name, our framework designed to enforce cryptographic \ac{mudb}.
Sec.~\ref{sec:reqs} defines the design requirements for \name.
Sec.~\ref {sec:MMM} introduces the concept of \ac{mmm}, and Sec.~\ref{sec:KeyManagement} presents our proposed \ac{mud} file extension to enable the MUD Controller to retrieve the necessary cryptographic material to carry out \ac{mudb} verification.
Then, in Sec.~\ref{sec:BindingAttestation} we detail \name\ \ac{zkp}-based verification procedure, including parameter embedding and profile update.
Table~\ref{tab:Symbols} summarizes the terminology used in this work.

\input{Tables/Symbols}

\input{Tables/Requirements}
\subsection{Design Requirements} \label{sec:reqs}
\name\ relies on five design requirements, summarized in Table~\ref{tab:Requirements}.
Although \acp{pki} provide a natural way to bind identities and resources, setting up and maintaining such an infrastructure introduces considerable complexity and management costs~\cite{LightweightPKI, PKIiot}.
Therefore, although an extension for MUD based on X.509 is available for secure URL issuance~\cite{MUD-rfc}, current \ac{mud} implementations rather rely on the insecure \ac{dhcp} extension~\cite{DefBehavior, NistMUD, CiscoMUD, SoftMUD, osMUD, SoKMUD, EnforcingBehavioral}.
These considerations motivate our design requirement of \emph{avoiding device-certificate \ac{pki} for \ac{mudb}} (\textbf{R1}), and rather rely on the plug-and-play \ac{dhcp} extension. 
This consideration creates a deployment-driven need to secure the \ac{dhcp} extension for URL issuance.
However, several security and implementation challenges arise when establishing secure \ac{dhcp}-based \ac{mudb} against our advanced threat model (Sec.~\ref{sec:SysModel}) without relying on a \ac{pki} infrastructure.
A major limitation of the solutions proposed in the literature is their lack of standard compliance.
Adopting non-standard compliant solutions significantly reduces interoperability and increases implementation complexity.
This particularly holds for the Controller, which would need to support multiple non-standard URL issuance mechanisms.
Therefore, any security mechanism and \ac{mud} extension should fit the standard by design, seamlessly integrating with the existing specification to increase interoperability and avoid increasing complexity.
This directly motivates the need for \emph{standards compliance and backward compatibility} (\textbf{R2}), ensuring that the proposed solution can be incrementally deployed without disrupting existing infrastructures.
Another challenge arises when considering symmetric-key approaches.
At first glance, mechanisms based on shared secrets (e.g., HMAC) appear attractive due to their efficiency.
However, the \ac{mud} Controller would need access to device-specific secret keys, which require secure key provisioning from manufacturers.
Besides increasing operational costs due to secure key provisioning, such a strategy introduces critical security issues.
An attacker could legitimately set up a malicious Controller to recover secret keys of target devices and reuse them to impersonate other devices.
Preventing such misuse leads to the requirement of \emph{resistance to \ac{mud} URL redirection and impersonation attacks} (\textbf{R3}).
Moreover, an overlooked implementation aspect in the literature involves the role of manufacturers in the \ac{mud} ecosystem.
Existing approaches heavily rely on manufacturer-assisted authentication, where devices or controllers interact with manufacturer-operated services at runtime.
However, manufacturers of \ac{iot} devices typically produce and maintain tens or even hundreds of products.
Thus, it would be extremely expensive for such manufacturers to maintain an additional always-on infrastructures that actively responds to authentication requests at every device deployment.
We argue that such operational costs would ultimately discourage the adoption of \ac{mud} in their products. 
Therefore, a key challenge is to design a solution that enforces secure \ac{mudb} without requiring active manufacturer participation, motivating the requirement of \emph{minimizing manufacturer involvement} (\textbf{R4}) and limiting them to hosting and maintaining information on a public web page rather than actively interacting with online parties.
Finally, a further challenge lies in the dynamic nature of \ac{iot} devices.
\ac{mud} profiles could change over the life of a device, e.g., due to a software update, firmware upgrade or inadvertent mistakes~\cite{DefBehavior, AcceptableURL}.
For example, a smart camera may receive a firmware update requiring it to contact new vendor analytics servers that were not part of its original profile.
Ensuring that updates are correctly propagated and securely enforced considering aforementioned challenges is a non-trivial problem.
This leads to the requirement of \emph{supporting secure \ac{mud} profile updates} (\textbf{R5}).

\subsection{\ac{mud} Management Model}  \label{sec:MMM}
An efficient \ac{mud} management strategy is essential to balance security, system complexity, updates, and overhead related to keys, \ac{mud} files, and certificate storage.
Despite its importance, existing works focus on the technical implementation of \ac{mud}-based solutions, overlooking how manufacturers are expected to manage their \ac{mud} files in practice.
Assigning a unique \ac{mud} file and \ac{url} to each device, although technically feasible, is inefficient.
Devices with the same network behavior would unnecessarily be associated with distinct files and URLs, resulting in an increased management and signing overhead.
At the same time, we observe that \ac{mud} files tend to be oriented toward classes of devices rather than individual instances~\cite{MUD-rfc, MUDCoAP}.
Building on this observation, we propose a \ac{mmm} in which devices are grouped into classes, each assigned a single \ac{mud} profile.
Devices within the same class share identical network requirements and are therefore linked to the same \ac{mud} file and \ac{url}.
For instance, a smart camera and a smart doorbell from the same manufacturer, although having different purposes, may require the same network services, thus having the same network profile.
This model significantly reduces management and storage overhead, as the number of \ac{mud} files to sign and store scales with the number of distinct MUD profiles, rather than with the total number of \ac{iot} devices.

\begin{definition}[MUD Class]
    A MUD Class is a group of devices that share the same network requirements for proper functioning.
    Each MUD class corresponds to a single MUD profile and is associated with one \ac{mud} file.
    A MUD class \(MUD_C\), defined within a manufacturer \(M\), is uniquely identified by \(MUD_{ID} = \left[M, MUD_C\right]\).
\end{definition}

Manufacturers are uniquely identified using standardized identifiers, such as the \ac{oui} or \ac{eui}~\cite{ieee2018guidelines}.
Different firmware versions may result in different network profiles, leading to distinct \ac{mud} files
classes.
Conversely, different firmware versions that however exhibit the same network behavior belong to the same MUD class.

\textbf{\ac{mud} File Update}.
During a device’s operational lifecycle, software updates may alter its network behavior~\cite{DefBehavior, AcceptableURL}, thereby requiring a change in the associated MUD profile and reassignment to a different MUD class.
In this case, the device must receive the \ac{mud} \ac{url} and necessary cryptographic material as part of the update content. 
This triggers a \ac{url} reissuance process, in which the MUD Controller obtains the new \ac{mud} file and updates the device’s network policy accordingly.
Additionally, \ac{mud} files may occasionally contain errors or incomplete information, which requires corrections or replacements~\cite{AcceptableURL}.
In this second case, the device and the MUD Controller are typically unaware of the change~\cite{AcceptableURL}.
To address this, we adopt a publish-subscribe model~\cite{SoKMUD}: the MUD Controller stores the signature of the current \ac{mud} file and periodically checks for updates.
A change in the signature indicates a file update, triggering the MUD Controller to retrieve and validate the new file, updating network policies accordingly.

\subsection{Key Management and \ac{mud} File Extension} \label{sec:KeyManagement}
We propose a key management approach that associates a secret key \(K_c\) with each \ac{mud} class \(C\).
Every manufacturer provides \ac{iot} devices with the secret key corresponding to their \ac{mud} class, namely $K_c$.
According to our system model, we consider this key to be stored in secure hardware, ensuring protection against physical extraction.
For each class, we also define public cryptographic parameters: an \ac{ec}, a base point \(G\) of prime order \(n\), and the point \(X_c = K_c \cdot G\) on the curve.
The point \(X_c\) binds the secret key \(K_c\) to the MUD class \(C\), enabling verifiable association without exposing the secret.
We thus propose extending the standard \ac{mud} file template with a cryptographic module containing these parameters together with \(MUD_{ID}\).
Listing~\ref{lst:MUDextension} provides the extended \ac{mud} file template publishing the above parmeters.
In this extension, \texttt{manufacturer-id} and \texttt{class-id} represent the manufacturer identifier and the \ac{mud} class identifier, respectively; \texttt{curve} specifies the elliptic curve name, which determines the base point \(G\); \texttt{p-format} indicates whether the elliptic curve point \(X_c\) is given in compressed or uncompressed format; \texttt{class-binding} contains the public point \(X_c\); \texttt{hash-alg} defines the hashing algorithm used during \ac{mudb} verification.
Upon signature verification, the Controller extracts this cryptographic information to carry out the \ac{mudb} procedure.
This approach is efficient since it leverages the \ac{mud} class abstraction, reducing the number of keys to manage from per-device to per-class.
\begin{lstlisting}[language={json}, label={lst:MUDextension}, caption={MUD file extension for \name\ setup}]
<-"ietf-mud:mud": { ->
<-"mud-version": 1, ->
<-"mud-url": "<URL>", ->
<-"last-update": "<YY-MM-DD>T<HH:MM:SS>Z", ->
<-"cache-validity": <HOURS>, ->
<-"is-supported": true, ->
<-"systeminfo": "<DESCRIPTION>", ->

<<"crypto-fidem:crypto": {>>
  <<"mud-identity": {>>
    <<"manufacturer-id": "<HEX>",>>
    <<"class-id": "<HEX>">>
  <<},>>
  <<"ec-params": {>>
    <<"curve": "<CURVE-NAME>",>>       
    <<"p-format": "<UNCOMPRESSED/
                    COMPRESSED>",>>
    <<"class-binding": "<EC-POINT>">>
  <<},>>
  <<"hash-alg": "<sha-256>">>
<<},>>

<-"from-device-policy": { ->
<-  ... ->
<-}, ->
<-"to-device-policy": { ->
<-  ... ->
<-} ->
<-}, ->

<-"ietf-access-control-list:acls": { ->
<-... ->
<-} ->
\end{lstlisting}

\subsection{MUD Binding Verification} \label{sec:BindingAttestation}
\textbf{Binding Verification Problem}.
Ensuring \ac{mudb} under the defined design constraints and \ac{mmm} requires an \ac{iot} device to prove to the Controller that it possesses the secret key \(K_c\) associated with the public value \(X_c\) specified in the MUD file.
At the same time, such a proof must not reveal \(K_c\) to the Controller, and neither require its direct disclosure or transfer.
This requirement is fundamental to prevent key leakage, preserve security under the considered threat model, and avoid the limitations of symmetric-key approaches.
\ac{zkp} provides a well-suited cryptographic primitive to address this mentioned challenge, allowing us to design a scheme to enforce \ac{mudb} while satisfying the constraints of standard compliance, scalability, and minimal manufacturer involvement.

\begin{figure}[!t]
    \centering
    \includegraphics[width=\columnwidth]{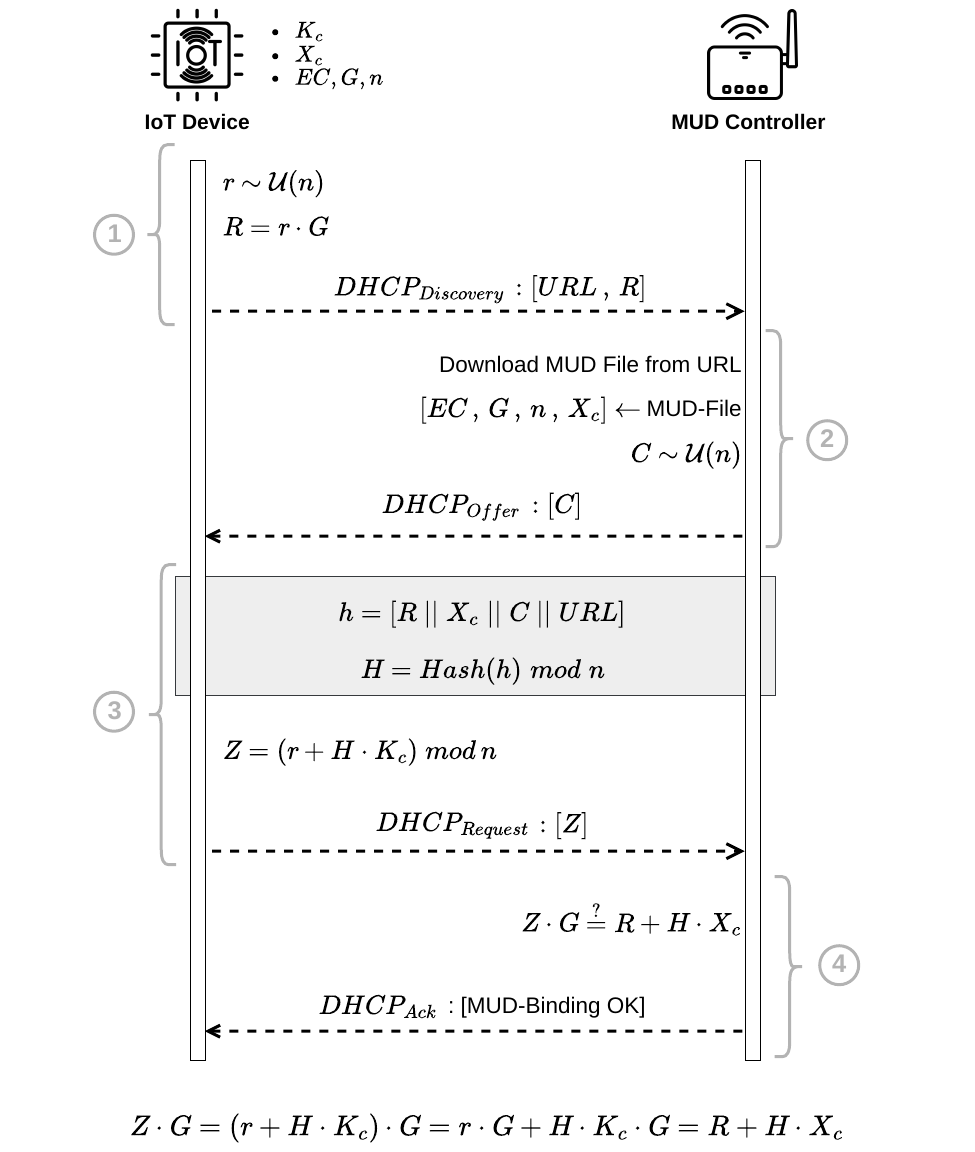}
    \caption{\name\ \ac{zkp}-based MUD Binding verification procedure avoiding manufacturers participation.}
    \label{fig:BindingAttestation}
\end{figure}

\textbf{Binding Verification Mechanism}.
During manufacturing, the \ac{iot} device is equipped with the \ac{mud} \ac{url} associated to the MUD class it belongs to, the corresponding secret key \(K_c\), and the MUD class public parameters \(X_c\), \(EC\), \(G\), \(n\).
At network joining time, the \ac{iot} device has to securely deliver its \ac{mud} \ac{url}, proving the \ac{mudb} property.
To this end, it performs the \ac{mudb} verification procedure shown in Fig.~\ref{fig:BindingAttestation} and discussed below.
\begin{enumerate}
    \item The \ac{iot} device samples a random value \(r \sim \mathcal{U}\left(n\right)\) and stores it in secure memory.
    Then, it computes the EC point \(R = r \cdot G\) as commitment for the sampled \(r\).
    It then sends a \ac{dhcp} \textit{Discovery} message including option 161 (see Sec.~\ref{sec:Embedding}), which carries the \ac{mud} \ac{url} and the commitment \(R\).

    \item The MUD Controller extracts the \ac{url}, retrieves the referenced \ac{mud} file, and verifies its authenticity.
    Upon verification, it retrieves the curve name \(EC\) with the corresponding parameters \(G\), \(n\), and the key commitment \(X_c\).
    It then samples a random challenge \(C\) and includes it in option 161 of the \ac{dhcp} \textit{Offer} message.

    \item The \ac{iot} device computes the \ac{zkp} response as \(Z = r + H \cdot K_c\), where \(H = Hash\left(R \; || \; X_c \; || \; C \; || \; URL\right)\), and \(Hash(\cdot)\) is a secure hash function~\cite{hash-review}.
    The device includes \(Z\) in option 161 of the \ac{dhcp} \textit{Request} message.
    In parallel, the MUD Controller computes the expected hash value \(H\) as well.

    \item The MUD Controller extracts the received \ac{zkp} response, and verifies the equality \(Z \cdot G \stackrel{?}{=} R + H \cdot X_c\).    
    If the check passes, the \ac{mudb} verification is successful.
    In such a case, the MUD Controller responds with a \ac{dhcp} \textit{Ack} message, translates the \acp{acl} in the \ac{mud} file into network polices, and enforces them.
    Otherwise, the procedure is aborted and network access is denied. 
\end{enumerate}

Upon device authentication, the MUD Controller saves a table, termed as \textit{MUD table}, with entries the device MAC address, its \(MUD_{ID}\), the received \ac{mud} \ac{url}, and the \ac{mud} file signature.

\subsubsection{\ac{zkp} Parameters Embedding} \label{sec:Embedding}
\begin{figure}[!t]
    \centering
    
        \begin{subfigure}[t]{0.75\linewidth}
            \centering
            \includegraphics[width=\linewidth]{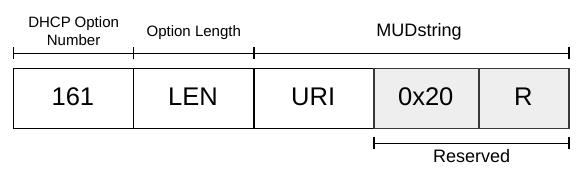}
            \caption{Embedding of commitment into the \texttt{reserved} space of the \texttt{MUDstring}.}
            \label{fig:Embedding1}
        \end{subfigure}
        \hfill
        
        \begin{subfigure}[t]{0.75\linewidth}
            \centering
            \includegraphics[width=\linewidth]{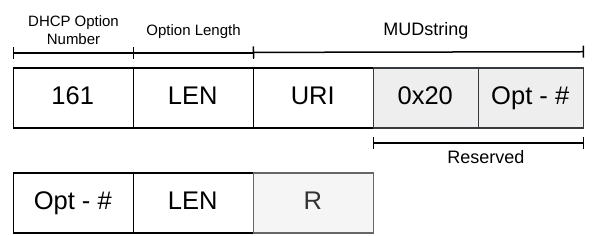}
            \caption{Embedding \(R\) into an available option. The \texttt{reserved} string contains the number of the options used.}
            \label{fig:Embedding2}
        \end{subfigure}
        
    \caption{\name\ \ac{zkp} parameters embedding in \textit{\ac{dhcp} Discovery} message.}
    \label{fig:Embedding}
\end{figure}

As backward compatibility is one of our priorities and constraints, it is essential to preserve message formats as defined by the standard.
To achieve this, we leverage the available \texttt{reserved} field in the \texttt{MUDstring} to carry the commitment \(R\) as illustrated in Fig.~\ref{fig:Embedding1}~\cite{MUD-rfc}.
However, the space available in the \texttt{reserved} field may not suffice to include the full \ac{ec} point.
In fact, \ac{dhcp} messages impose strict size limitations: each option can carry at most 255 bytes, and the entire \ac{dhcp} message cannot exceed 576 bytes~\cite{DHCP}.
These constraints can make it challenging to embed all required data within a single option.
To address this limitation, we propose an alternative approach: include the commitment \(R\) in a separate available \ac{dhcp} option and specify the option number used in the \texttt{reserved} field, as shown in Fig.~\ref{fig:Embedding2}.
This design ensures interoperability with devices from different vendors, which may already use custom \ac{dhcp} options for other purposes.
By providing flexibility in option allocation, we avoid collisions and maintain compatibility across heterogeneous environments.

\subsubsection{\ac{mud} Profile Update} \label{sec:ProfileUpdate}
Maintaining correct and updated network restrictions on \ac{mud}-enabled \ac{iot} networks requires a secure \ac{mud} profile update mechanism.
Fig.~\ref{fig:mud_update_flow} illustrates the flow chart of the file update logic of \name.  
To address potential changes in \ac{mud} files due to corrections, \name\ leverages a publish-subscribe approach~\cite{SoKMUD}.
The \ac{mud} Controller periodically verifies whether the \ac{mud} file signatures of connected devices differ from those stored in the \ac{mud} table.
If a mismatch is detected, the Controller automatically retrieves the updated \ac{mud} file, refreshes its signature, and enforces the new network restrictions.
Conversely, when an \ac{iot} device already registered in the \ac{mud} table issues a new \ac{url}, the \ac{mudb} verification procedure must be executed again.
In this case, an additional check is required compared to the initial registration.
We observe that a device cannot change its manufacturer during its lifecycle~\cite{AcceptableURL}.
Therefore, if the new \ac{url} points to a \ac{mud} file associated with a different manufacturer, the MUD Controller aborts the procedure, and if possible, alerts the network administrator of suspicious behavior.
In scenarios where a software update does not alter the device profile, the device typically reboots to apply the update.
This reboot triggers a new \ac{dhcp} handshake.
Since the profile remains unchanged, the \ac{url} matches the entry stored in the \ac{mud} table.
In this specific case, the Controller can skip the verification procedure, and omit the challenge \(C\) in the \ac{dhcp} \textit{Offer} message.


%% file: Tables/Symbols.tex
\begin{table}[!h]
\centering
\caption{Notation used in this work.}
\label{tab:Symbols}
\resizebox{.95\columnwidth}{!}{%
\begin{tabular}{c|l}
    \hline
    \rowcolor[HTML]{EFEFEF} 
    \textbf{Notation} & \multicolumn{1}{c}{\cellcolor[HTML]{EFEFEF}\textbf{Description}} \\ \hline

    \(M\)     & Manufacturer identifier \\ 
    \(MUD_C\)   & MUD Class identifier   \\
    \(MUD_{ID}\) & MUD identifier: [\(M\) , \(C\)] \\
    \(EC\)       & Elliptic-Curve                                        \\
    \(G\)        & Base point of the Elliptic-Curve of order \(n\)       \\
    \(n\)        & Order of the Elliptic-Curve                           \\
    \(K_c\)      & Secret key associated to the MUD Class \(C\)     \\
    \(X_c\)      & Public commitment of \(K_c\) for the MUD Class \(C\) \\
    \(Hash(\cdot)\) & Secure hash function \\
    \(H\) & Hash value output from \(Hash(\cdot)\) \\
    \(r\)   & Random high-entropy nonce for \ac{zkp} \\
    \(R\) & Commitment of random nonce \(r\) for \ac{zkp} \\
    \(C\) & Random challenge issued by the MUD Controller \\
    \(Z\) & \ac{zkp} response for \ac{mudb} verification

    \end{tabular}%
}
\end{table}

%% file: Tables/Requirements.tex
\begin{table}[!t]
\centering
\renewcommand{\arraystretch}{1.25}
\caption{\name\ design requirements.}
\label{tab:Requirements}
\resizebox{0.95\columnwidth}{!}{%
\begin{tabular}{c|l}
    \hline
    \rowcolor[HTML]{EFEFEF} 
    \textbf{\(\#\)} & \multicolumn{1}{c}{\cellcolor[HTML]{EFEFEF}\textbf{Design Requirements}} \\ \hline
    \textbf{R1} & Avoid device-side \ac{pki} dependence for \ac{mudb}\\
    \textbf{R2} & Standard compliance and backward compatibility \\
    \textbf{R3} & Prevent \ac{mud} \ac{url} redirection attacks   \\    
    \textbf{R4} & Minimize manufacturer involvement\\    
    \textbf{R5} & Enable secure \ac{mud} profile updates  \\
    
\end{tabular}%
}
\end{table}

%% file: Sections/6-SecurityAnalysis_modified.tex
\section{Security Analysis} \label{sec:SecurityAnalysis}
This section presents the security analysis of the \ac{mudb} verification procedure proposed in \name\, explaining how attacker objectives \textbf{O1–O3} are mitigated.

\subsection{Security Considerations}
\textbf{Security Foundations}.
The security of the \ac{mudb} verification mechanism presented in Sec.~\ref{sec:BindingAttestation} relies on two fundamental considerations.
First, each device stores the private key \(K_c\) in tamper-resistant memory, making physical extraction infeasible, in line with assumptions in prior works~\cite{MUDscan, hardware1, hardware3, hardware4}.
Moreover, recovering \(K_c\) from \(X_c\) is computationally infeasible under the \ac{ecdl} assumption~\cite{EClogarithm}.
The same holds for the nonce \(r\) with respect to the commitment \(R\).
Under these assumptions, an adversary observing the pair (\(R\), \(Z\)), and knowing the \ac{ec} parameters cannot recover \(K_c\) or \(r\).
Consequently, a malicious \ac{iot} device unaware of the correct \(K_c\) cannot compute a valid response to the challenge of the MUD Controller.
Since the \(K_c\) and \(r\) are never transmitted, these properties address the attacker objective \textbf{O1} (secret key recovery).   

\begin{figure}[!t]
    \centering
    \includegraphics[width=.8\columnwidth]{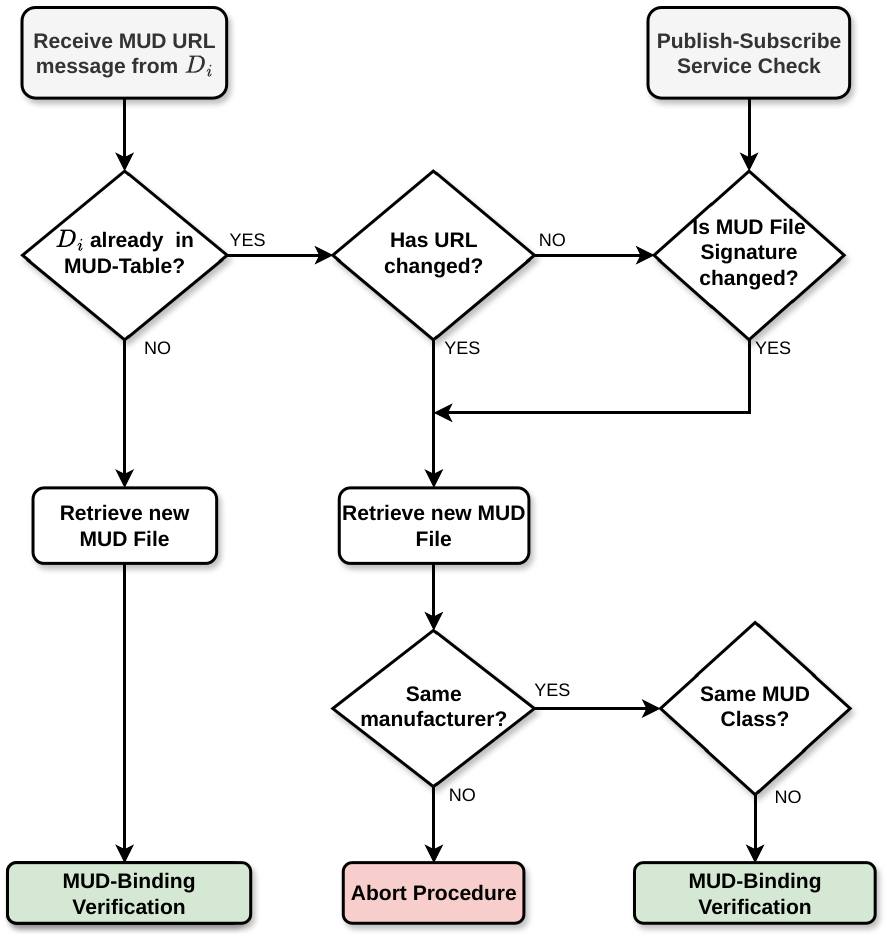}
    \caption{Flow chart of the MUD file update procedure logic implemented in FIDEM.    }
    \label{fig:mud_update_flow}
\end{figure}

\textbf{Manufacturer and Profile Integrity}.
\name\ enforces manufacturer and profile integrity through two mechanisms.
During initial verification, the MUD Controller validates the signature of the \ac{mud} file retrieved at the supplied \ac{url}, ensuring authenticity of the device manufacturer.
During profile updates, \name\ guarantees that a device cannot change manufacturer over its lifetime.
If a newly-issued \ac{mud} \ac{url} points to a file signed by a different manufacturer, the MUD Controller aborts the protocol and flags the event as suspicious.
This prevents compromised or counterfeit devices from migrating across manufacturers or issuing unauthorized \acp{url}.
Such properties address the attacker objective \textbf{O2} (cross-manufacturer impersonation).

\textbf{Resistance to Replay Attacks and Oracle Abuse}.
The attacker may record legitimate \((R', Z')\) pairs and replay them to bypass verification, luring the MUD Controller to enforce incorrect restrictions.
Alternatively, the attacker may leverage \(D_{oracle}\) as a cryptographic oracle to compute \ac{zkp} values on behalf of the compromised device \(D_1\).
To prevent such attacks, \name\ relies on an interactive Schnorr protocol in which each execution uses a fresh challenge \(C\).
Since a previously recorded \ac{zkp} response \(Z'\) is bound to an old challenge \(C'\) sampled by the MUD Controller, it cannot be replayed in a new session.
In fact, \(H' = Hash\left(R' \; || \; X_c \; || \; C' \; || \; URL\right)\) used to compute \(Z'\), differs from \(H = Hash\left(R' \; || \; X_c \; || \; C \; || \; URL\right)\) since \(C \neq C'\).
Furthermore, the \ac{mudb} verification occurs during the \ac{dhcp} handshake, before the device obtains network access.
Therefore, a compromised device \(D_1\) cannot communicate with \(D_{oracle}\) during the verification phase.
These properties eliminate adaptive oracle attacks and address the attacker objective \textbf{O3} (\ac{mud} profile escalation).

\textbf{Security Under Nonce Reuse Scenarios}.
Nonce uniqueness is critical for interactive \ac{zkp}-based security.
If the same \(r\) is reused across two sessions with different challenges, an adversary can eliminate it and compute \(K_c = (Z_1 - Z_2)(H_1 - H_2)^{-1} \quad mod \; n\), breaking the binding mechanism.
Therefore, the MUD Controller must track past commitments \(R\) to detect reuse.
Conversely, using different nonces with the same challenge does not compromise security, as \(H\) incorporates \(R\), ensuring unlinkability.

In summary, within the system and threat-model assumptions defined in Sec.~\ref{sec:SysModel}, \name\ addresses objectives O1–O3, and ensures that the Controller can securely verify \ac{mudb} even in the presence of compromised or counterfeit devices. 

\subsection{Security Assessment with ProVerif}
\begin{figure}[!t]
    \centering
    \includegraphics[width=\columnwidth]{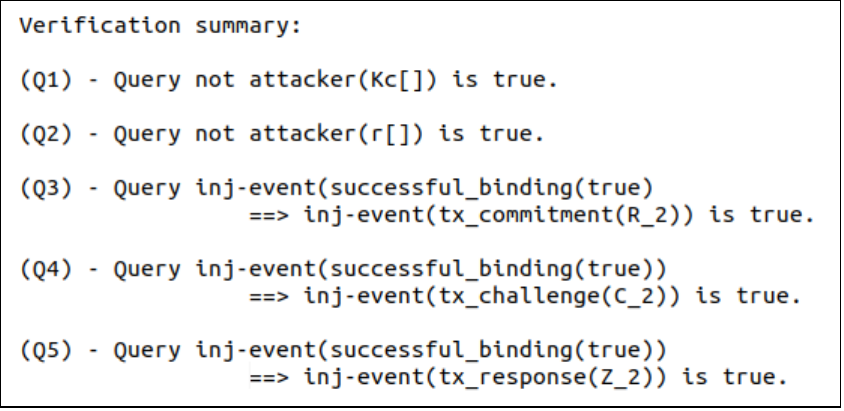}
    \caption{Output for \name\ formal verification with ProVerif. All queries are validated, proving \name\ security.}
    \label{fig:Proverif}
\end{figure}

To rigorously validate the security of the proposed \ac{zkp}-based binding mechanism (Sec.~\ref{sec:BindingAttestation}), we use ProVerif~\cite{Proverif}, a widely recognized tool for automated protocol analysis~\cite{ePPTM, SEC-004, A2RID, ARID}.
ProVerif is especially useful when creating new protocols that leverage existing cryptographic primitives, whose security has already been proven, as is the case for \name.
ProVerif evaluates the security of the protocol under the Dolev-Yao attacker model~\cite{SEC-004}.
If an attack is possible, ProVerif provides a trace to reproduce it.
We formally modeled \name\ in a single protocol round and executed multiple queries to verify secrecy, resistance to guessing attacks, and authentication properties.
Interested readers can find the source code of \name\ within ProVerif in our GitHub repository.
Fig.~\ref{fig:Proverif} summarizes the results of our formal security analysis.

The queries output Q1 and Q2 \textit{not attacker(<elem[])> is true} establish within the symbolic model the attacker cannot retrieve the value of \textit{elem[]}.
Otherwise, if the output is \textit{false}, the attacker can retrieve it.
They represent \ac{mud} class key \(K_c\) and the secret nonce \(r\), respectively.
Queries \textbf{Q3}, \textbf{Q4}, \textbf{Q5} \textit{inj-event(event\_A) \(==>\) inj-event(event\_B)} expresses an injective correspondence property.
It means that for every occurrence of event \textit{event\_A}, there is a unique corresponding occurrence of event \textit{event\_B} earlier in the protocol execution.
If the query output is \textit{true}, every successful \textit{event\_A} is guaranteed to correspond to a unique prior \textit{event\_B}, thus preventing replay attacks and ensuring that protocol steps are correctly linked.
In contrast, \textit{false} output implies the protocol cannot guarantee that every successful event\_A is tied to a unique prior event\_B, which undermines correctness and freshness guarantees.
In our ProVerif code, event \textit{successful\_binding(true)} represents successful \ac{mudb} verification, \textit{tx\_commitment(R)}, \textit{tx\_challenge(C)} and \textit{tx\_response(Z)} represent events of sending commitment \(R\), challenge \(C\), and \ac{zkp} response \(Z\), respectively.
Successful queries \textbf{Q3}, \textbf{Q4}, \textbf{Q5} thus confirm injective correspondence between successful bindings, commitments \(R\), challenges \(C\), and responses \(Z\), respectively, thereby preventing replay attacks and ensuring freshness.
In summary, our automated analysis through ProVerif shows that \name\ provides strong secrecy and \ac{mudb} guarantees.


%% file: Sections/7-ExperimentalEvaluation.tex
\section{Experimental Evaluation} \label{sec:Evaluation}
This section presents our extensive experimental evaluation of \name.
\name\ \ac{poc} source code is available on GitHub.

\subsection{Experimental Setup and Settings} \label{sec:exp_setup}
\begin{figure}[!t]
    \centering
    \includegraphics[width=\linewidth]{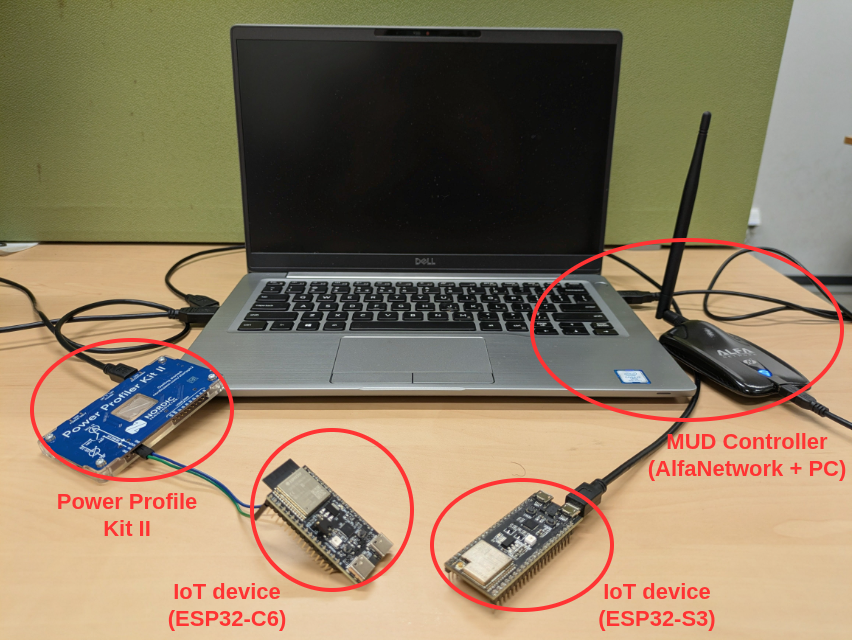}
    \caption{\name\ \ac{poc} implementation setup.}
    \label{fig:Setup}
\end{figure}

\textbf{Setup}.
Fig.~\ref{fig:Setup} illustrates the setup of our experimental \ac{poc} implementation of \name.
We use two reference \ac{iot} devices.
We first consider the ESP32-S3, an embedded resource-constrained \ac{iot} device
equipped with a dual-core XTensa LX7 MCU, 512 KB of internal SRAM, and integrated with a 2.4 GHz, 802.11 b/g/n WiFi card.~\footnote{\url{https://www.espressif.com/en/products/socs/esp32-s3}}
This device does not feature secure storage capabilities, allowing us to deploy and test our solution on a general-purpose hardware not equipped with advanced security features.
Moreover, we also consider the ESP32-C6 microcontroller, equipped with a 32-bit RISC-V processor, a 320 KB ROM and a 512 KB SRAM, and integrated with a 2.4 GHz, 802.11 b/g/n WiFi card.~\footnote{\url{https://www.espressif.com/en/products/socs/esp32-c6}}
It also features a hardware \ac{tee}, providing {\em secure storage} and protected cryptographic operations.
Including such a device in our \ac{poc} and testing performance on it allows us to evaluate the impact on performance of a hardware capable of enforcing protection of cryptographic secrets stored onboard the device.
We implement the MUD Controller using a DELL Latitude 7400 laptop equipped with Ubuntu 22.04.5 LTS and an AlfaNetwork antenna.~\footnote{\url{https://www.alfa.com.tw/products/awus036nha?_pos=6&_sid=569e0ca77&_ss=r&variant=36473966166088}}
We collect energy consumption using a Power Profiler Kit II from Nordic,~\footnote{\url{nordicsemi.com/startppk2}} a high-precision measurement tool designed to monitor real-time current draw and energy usage of embedded devices.
It connects directly to the \ac{iot} device and provides accurate profiling of power consumption during different operational phases.
For implementation, we use the standard \ac{ec} \emph{secp256r1}~\cite{secp256}.

\textbf{Settings}.
We experimentally evaluate \name\ with respect to two key metrics: (i) computational overhead and (ii) additional energy consumption induced by the \ac{zkp}-based \ac{mudb} verification procedure.
For comparison, we consider three benchmarks: (i) \textit{standard \ac{dhcp}}, following the standard RFC8520~\cite{MUD-rfc}; (ii) \textit{X.509 with HTTP}, using the standard X.509 certificate extension, shared via HTTP payload; and (iii) \textit{X.509 with TLS}, using the standard X.509 certificate extension, used for TLS handshake~\cite{TLS}.
Where possible, we implement \name\ with and without using the \ac{tee}, to assess the overhead due to \ac{tee} involvement and computations.
For each scenario, we collect data from 100 protocol runs and report results through boxplots, showing the median value of the metric of interest.
We assess the computational overhead by measuring the time required to carry out the \ac{mudb} verification procedure presented in Sec.~\ref{sec:BindingAttestation}.
Therefore, we consider the elapsed time of the overall \ac{mudb} verification procedure at MUD Controller side, from the reception of the \ac{dhcp} \textit{Discovery} message, up to reception and verification of the \ac{zkp} response received.
Note that this is the overall time to complete the procedure, including also operations executed on the \ac{iot} device under test.
We evaluate energy consumption only on the \ac{iot} device, as the MUD Controller is not resource-constrained and its energy usage is not a concern.
We consider the energy consumed starting from system setup (including cryptographic context initialization) up to the reception of the \ac{dhcp} \textit{Ack} message, signaling successful \ac{mudb}.

\subsection{PoC Implementation Details}    \label{app:PoCimplementation}
For \ac{poc} implementation and overhead analysis, we omit procedures and communications that do not vary across \ac{mudb} verification approaches.
Remote \ac{mud} file retrieval, for example, is common to all considered approaches and is therefore excluded.
Instead, \ac{mud} files are stored locally, but file certificate verification remains implemented.
Each file is signed using a locally generated private key that emulates the manufacturer's signing key, while the corresponding public key is certified by a locally generated certificate authority.
We use the OpenSSL API for key generation and cryptographic operations.

\textbf{MUD Controller Implementation}
The \ac{mud} controller runs on an Ubuntu laptop equipped with an Alfa Network AR9271 USB adapter configured as an access point.
We use \texttt{hostapd} and \texttt{systemd-networkd} to control the network interface and configure the access point with a static IP address.
A Kea DHCPv4 server manages DHCP exchanges and exposes hooks for integrating custom logic and handling DHCP options.
Specifically, we implement the \texttt{pkt4\_receive} and \texttt{pkt4\_send} callouts to inspect incoming and outgoing packets and log custom options, including Option~161 carrying the MUD URL.

For the X.509-based approaches, we use OpenSSL for cryptographic computations and implement the X.509 extensions required for MUD URL issuance via HTTP or a TLS handshake.
At startup, the controller launches a dedicated thread that listens on the configured port and handles certificate exchange and MUD URL transmission.
We measure \ac{mudb} verification time at the controller using the POSIX \texttt{clock\_gettime()} function.

\textbf{IoT Device Implementation}
The \name\ device-side logic runs on an ESP32 microcontroller and uses the Espressif API, whose DHCP hooks enable packet manipulation analogous to the controller-side Kea callouts
We integrate basic HTTP and TLS clients to transmit X.509 certificates and share the MUD URL with the controller.
Cryptographic operations use hardware acceleration and execute in an asynchronous worker thread.
Upon receiving the challenge, the worker computes the hash value \(H\) and the \ac{zkp} response \(Z\), while the main task concurrently constructs the standard DHCP Request message before invoking the transmission hook.
By overlapping these operations, the implementation minimizes the latency added by the cryptographic computations.

\textbf{FIDEM Implementation}
\name's binding verification uses the elliptic curve \texttt{secp256r1}, which is supported by OpenSSL.
The nonce \(r\), challenge \(C\), and secret key \(K_c\) are all 256 bits cryptographic parameters.
Because DHCP options have limited payload capacity, \name\ carries its parameters in custom DHCP options. Specifically \(R\) is embedded as shown in Fig.~\ref{fig:Embedding2}.

\subsection{Results} \label{sec:results}
\begin{figure*}[!t]
    \centering
    
        \begin{subfigure}[t]{\columnwidth}
            \centering
            \includegraphics[width=\columnwidth]{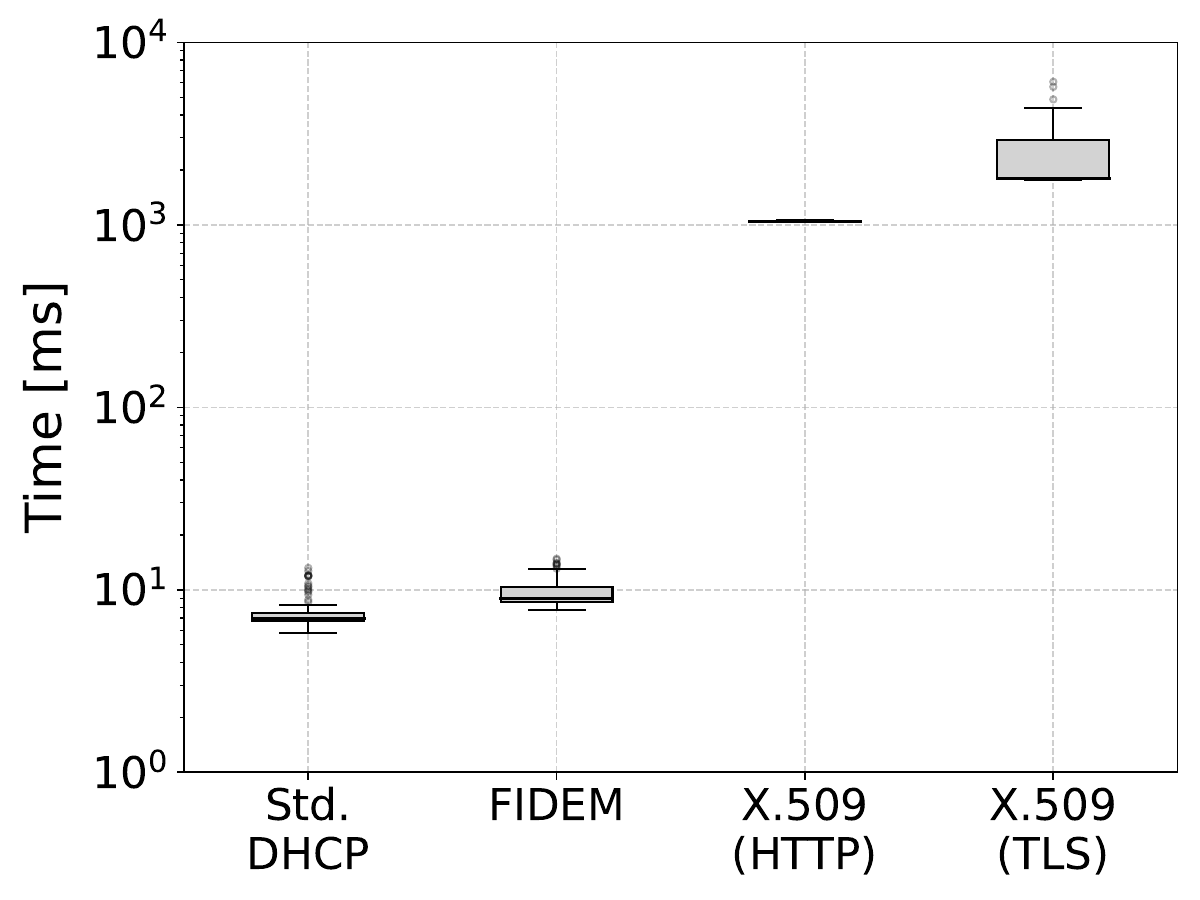}
            \caption{Verification time with ESP32-S3 (Controller side).}
            \label{fig:S3-TimeOverhead}
        \end{subfigure}
        \hfill
        \begin{subfigure}[t]{\columnwidth}
            \centering
            \includegraphics[width=\columnwidth]{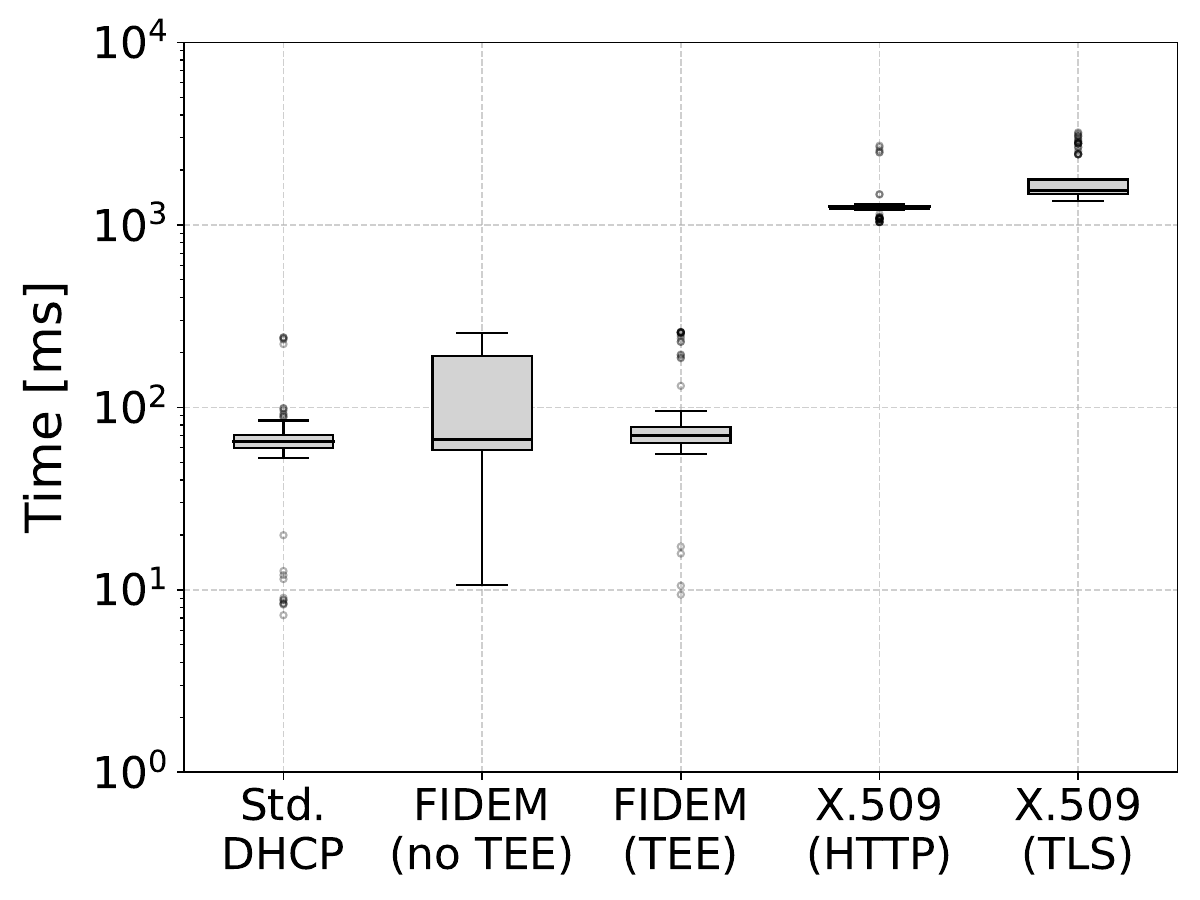}
            \caption{MUD-Binding with ESP32-C6 (Controller side).}
            \label{fig:C6-TimeOverhead}
        \end{subfigure}
        
        \begin{subfigure}[t]{\columnwidth}
            \centering
            \includegraphics[width=\columnwidth]{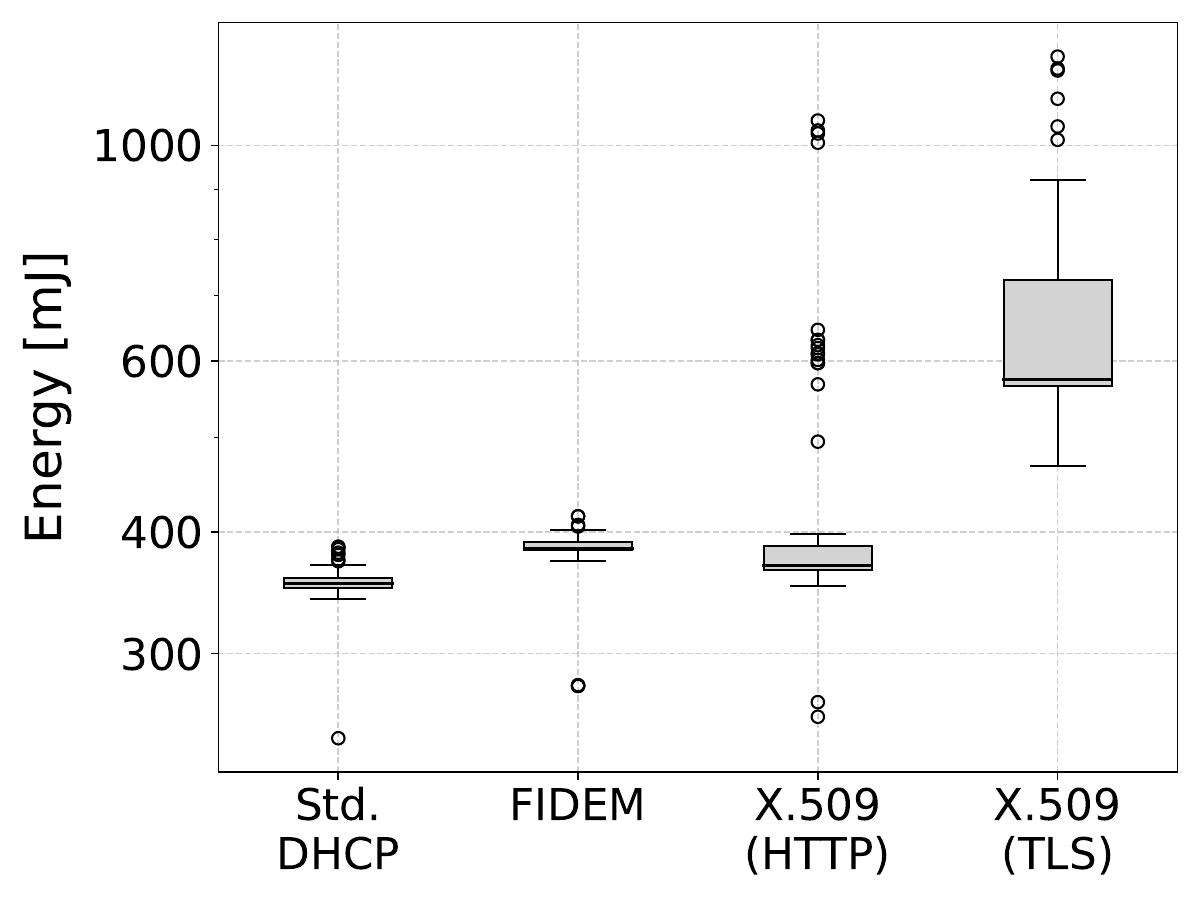}
            \caption{ESP32-S3 verification energy (IoT side).}
            \label{fig:S3-EnergyOverhead}
        \end{subfigure}      
        \hfill
        \begin{subfigure}[t]{\columnwidth}
            \centering
            \includegraphics[width=\columnwidth]{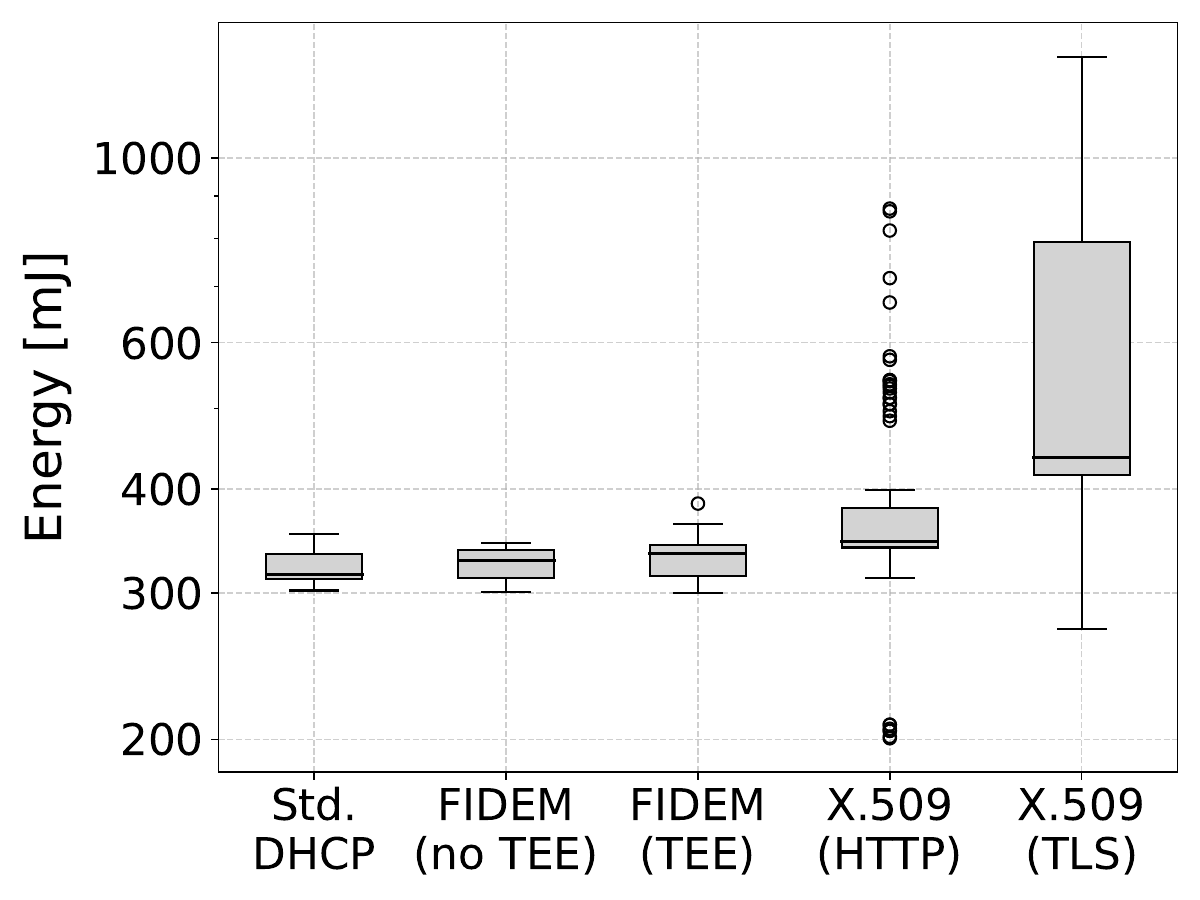}
            \caption{ESP32-C6 verification energy (IoT side).}
            \label{fig:C6-EnergyOverhead}
        \end{subfigure}

    \caption{MUD-Binding verification time and energy using \name.}
    \label{fig:ExpResults}
\end{figure*}

Fig.\ref{fig:ExpResults} reports our complete experimental results in terms of computational and energy overhead for \ac{mudb} verification implementing \name\ over ESP32-S3 (Fig.~\ref{fig:S3-TimeOverhead},~\ref{fig:S3-EnergyOverhead}) and ESP32-C6 (Fig.~\ref{fig:C6-TimeOverhead},~\ref{fig:C6-EnergyOverhead}) devices.

Considering the execution time with the ESP32-S3 device (Fig.~\ref{fig:S3-TimeOverhead}), \name\ introduces minimal delay compared to the standard DHCP method.
\name\ requires a median time of \(9.06ms\), which is \(\sim 30\%\) more than the time required by the standard \ac{dhcp} issuing method (\(6.99ms\)).
We recall that, although lightweight, the standard DHCP mechanism does not ensure \ac{mudb}, being thus unsuitable for secure deployments.
At the same time, \name\ significantly outperforms X.509 leveraging both HTTP and TLS, which require a median time of \(1.05~s\) and \(1.79~s\), respectively.
Such a large time difference is due to the additional operations required by connection establishment.
X.509 with HTTP requires setting up a TCP connection, and X.509 with TLS adds several further cryptographic \ac{pki}-based operations during the handshake to set up a secure connection.
We acknowledge that such a secure connection provides many more usable security guaranties beyond secure device enrollment and \ac{mudb}; however, it requires a \ac{pki}, expensive to deploy and manage for \ac{iot} manufacturers.

When considering the energy consumption with the ESP32-S3 (Fig.~\ref{fig:S3-EnergyOverhead}), we notice that standard DHCP method exhibits the lowest energy usage, i.e., a median of  \(354mJ\).
We can explain this result by recalling that no security checks are performed when using this mode.
Our proposed solution, FIDEM, introduces a slight increase compared to DHCP, i.e., a median consumption of \(385mJ\), for an overhead of $31mJ$ (\(\sim8,7\%\)). 
This negligible energy overhead demonstrates that \ac{ec}-based \ac{zkp} operations are suitable for constrained devices.
X.509 with HTTP shows an energy consumption comparable to \name\, i.e. a median of \(370 mJ\).
Here, we notice an apparent mismatch between the relationship of time and energy consumption between X.509 with HTTP and \name: although \name\ is quicker, it consumes approximately the same energy.
We can explain these results by recalling that \name\ relies on cryptographic operations executed in the \ac{caam} module of the ESP-32 device.
Activating and using such dedicated hardware requires additional energy, partly outweighing the energy gained from the reduced duration of the protocol compared to X.509 over HTTP.
Moreover, we recall that using HTTP to issue the X.509 certificate lacks security guarantees: the certificate is transmitted as a plain payload and can be replayed, failing to ensure the binding property (R3, see Sec.~\ref{sec:reqs}).
Conversely, X.509 over TLS incurs a substantial energy overhead, i.e., a median of \(574 mJ\), nearly $\times 1.5$ that of \name, with wide variance due to the complexity of the TLS handshake.
This high consumption is due to the time required to complete message exchange (as noticed for Fig.~\ref{fig:S3-TimeOverhead}) and due to the higher number of cryptography operations involved. Furthermore, during the tests, we observed several network failures, which have an additional influence on the time required to complete the \ac{mudb} process.

When experimentally evaluating \name\ on the ESP32-C6 device, we observe that the performance trend follows the one already discussed for the ESP32-S3 device: \name\ is about \(\times 21\) faster and consumes \(23\%\) less energy compared to X.509 over TLS.
\name\ (TEE) takes $70ms$ and $334mJ$, while X.509 over TLS takes $1535ms$ and $436mJ$.
At the same time, compared to the trends observed for the ESP32-S3 device, we notice that using the ESP32-C6 device increases the baseline time required for \ac{dhcp} handshake.
This is evident when comparing the standard DHCP data series between Fig.~\ref{fig:S3-TimeOverhead} and Fig.~\ref{fig:C6-TimeOverhead}.
This is due to several factors, including differences in architecture, hardware, and wifi cards between the two models, which lead to a larger fixed connection plus stack cost.
Nevertheless, the ESP32-C6 device is natively optimized for cryptographic operations, allowing to reduce \name\ (TEE) time overhead compared to standard DHCP to only about \(5 ms\) and energy consumption to \(20 mJ\).
Furthermore, we also compare the performance of \name\ on the ESP32-C6 device with and without using the \ac{tee} secure hardware, to evaluate the additional delay and energy consumption introduced by such a component.
Fig.~\ref{fig:C6-TimeOverhead} shows that leveraging the \ac{tee} introduces limited computational delay and increased energy(\(\sim 3.5 ms\) and \(\sim6 mJ\)).
This overhead is induced by the execution of required operations for switching from the insecure to the secure worlds.
Finally, we notice that although taking longer execution times, the ESP32-C6 requires less energy consumption compared to the ESP32-S3.
Specifically, \name\ (no TEE) requires $328mJ$ and \name\ (TEE) \(\sim334 mJ\).
This indicates a lower average power profile and design of the ESP32-C6, which favors idle states during waiting periods.

In summary, our results demonstrate that \name\ achieves strong security guarantees with negligible impact on the resources and lifetime of constrained \ac{iot} devices.

%% file: Sections/8-Extensions.tex
\section{Discussion} \label{sec:Limitations}
\textbf{Out-of-Band Communications}.
Aligned with the \ac{mud} standard RFC8520~\cite{MUD-rfc}, our system model considers only \ac{iot} traffic traversing a monitoring entity (e.g., a router), where \ac{mud} profiles are enforced.
Accordingly, out-of-band channels such as cellular networks or direct device-to-device communications (e.g., Bluetooth or ZigBee) are excluded.
While resource-constrained devices rarely support cellular connectivity, short-range communication technologies are commonly available in higher-end \ac{iot} devices.
These channels could be exploited by an attacker to bypass the router and offload cryptographic operations to \(D_{oracle}\), thereby undermining \ac{mudb} verification.
However, addressing such scenarios would require extending \ac{mud} beyond IP-based communications, entailing a fundamentally different architectural model.
This would significantly alter both the system and threat models assumptions considered in this work, effectively defining a distinct research problem rather than a direct extension of \name.
Investigating native \ac{mud} integration into non-IP protocol stacks therefore represents a promising direction for future work.

\textbf{Scalability}.
\name\ operates exclusively during the \ac{dhcp} handshake, i.e., before the device receives network connectivity and before any operational traffic is allowed.
As such, it does not introduce persistent computational or communication overhead during normal device operation.
In contrast, the only additional cost is a one-time, bounded delay during onboarding (\(\sim5 ms\)), as quantified in our evaluation.
Consequently, scalability in this context is related to the \ac{mud} Controller’s capability to handle multiple concurrent onboarding requests.
However, this is a property inherent to the implementation and deployment of the Controller, not to \name\ itself.
Similarly, potential \ac{dos} scenarios target the Controller’s exposure during the \ac{dhcp} phase.
However, this threat already exists in standard deployments, and \name\ does not amplify this attack surface beyond a minimal processing overhead.

\textbf{Extension to Non-Interactive Scenarios}.
Our design of \name\ focuses on the \ac{dhcp} extension due to its widespread adoption for IP address assignment and its support in current \ac{mud} implementations.
The \ac{mud} standard also defines an \ac{lldp} extension for URL issuance; however, \ac{lldp} is inherently non-interactive.
As a result, \name\ cannot be directly extended using non-interactive Schnorr-based \ac{zkp} authentication~\cite{Remote}, as this would expose the system to pre-computation and replay attacks.
A possible mitigation is to leverage interaction with the \textit{Operating Server} for device identity authentication.
However, device authentication alone does not guarantee \ac{mudb}: a malicious device may successfully authenticate while still issuing a spoofed URL.
To address this, after successful authentication, the Controller can rely on the verified device identity to directly retrieve the corresponding \ac{mud} file from the MUD Server, rather than trusting the URL advertised by the device.
While this approach involves the MUD Server during file retrieval, it does not require additional cryptographic operations.

\textbf{MUD Management Model and Group-Based Authentication.}
Our proposed \ac{mmm} associates a single \(K_c\) with each \ac{mud} class.
While we assume secure storage of cryptographic material on \ac{iot} devices, this design introduces an inherent limitation: if \(K_c\) is compromised, all devices within the corresponding \ac{mud} class are affected.
A potential mitigation, while still avoiding per-device certificates, is the adoption of group-based authentication schemes, where devices share a common public key but hold distinct private keys~\cite{SurveyGroupSignatures}.
In this setting, a class-level public key could be included in the \ac{mud} file, while each device is provided with a unique private key, thus overcoming the shared-secret limitation of our \ac{mmm}.
However, group signature schemes introduce additional challenges, including complex key revocation, scalability limitations, and dependence on trusted setup procedures~\cite{SurveyGroupSignatures, FoundationsGroupSignatures, IntegrityGroupSignatures}.
Considering such limitations and our system model, we believe that the proposed \ac{zkp}-based approach at the roots of \name\ offers the best compromise at present.
These trade-offs suggest that alternative \acp{mmm} require careful evaluation, motivating future research toward designs that better balance security, scalability, and efficiency.

%% file: Sections/9-Conclusions.tex
\section{Conclusions}   \label{sec:Conclusions}
In this paper, we presented \name, a framework designed to address the fundamental gap in the \ac{mud} standard by securing the widely adopted but insecure \ac{dhcp}-based \ac{mud} URL issuance mechanism.
\name\ leverages Schnorr-based \ac{zkp} authentication to cryptographically enforce the \ac{mudb} property while remaining fully standard-compliant and backward compatible.
By design, \name\ withstands a stronger adversarial model than existing approaches, including attackers capable of exploiting legitimate devices as cryptographic oracles.
Furthermore, it achieves this without requiring device certificates or device-side \ac{pki} for \ac{mudb}, avoiding the associated operational complexity, and without requiring active manufacturer involvement, thus preserving scalability and deployability in heterogeneous IP-connected, DHCP-based \ac{mud} environments.
Experimental results from a real-world \ac{poc} demonstrate that these security guarantees are achieved with minimal overhead compared to standard \ac{dhcp}, while significantly outperforming certificate-based solutions (\(\times 21\) faster and \(\sim23\%\) less energy consumption than X.509 over TLS on an ESP32-C6 device).
Future work will investigate extensions to LLDP-based deployments, mitigation of out-of-band communication threats, and alternative strategies such as group-based digital signatures.